\documentclass[journal]{IEEEtran}
\usepackage{xcolor}
\usepackage{fancyhdr}
\usepackage{tgschola}
\usepackage{lastpage}
\usepackage{amsthm,amssymb,amsmath,mathtools,fixmath}
\usepackage{amsbsy}
\usepackage{moreverb}
\usepackage{amsfonts}
\usepackage{graphicx}
\graphicspath{{SimulationFigures/}}
\usepackage{tikz}
\usepackage{listings}
\usepackage{color}
\usepackage{latexsym}
\usepackage{subfigure}
\usepackage{animate}
\colorlet{linkequation}{blue}
\usepackage[linesnumbered,ruled]{algorithm2e}
\usepackage{soul}
\DeclareMathOperator*{\argminA}{arg\,min} % Jan Hlavacek
\usepackage{lipsum}% http://ctan.org/pkg/lipsum
\usepackage[colorlinks,bookmarksopen,bookmarksnumbered,citecolor=black,urlcolor=black]{hyperref}
\hypersetup{linkcolor=black}
\DeclareMathOperator*{\maxi}{maximize}
\makeatletter
\def\BState{\State\hskip-\ALG@thistlm}
\makeatother

\ifCLASSINFOpdf
\else
\fi
\begin{document}
\title{\textcolor{black}{Load Balancing} User Association in Millimeter Wave MIMO Networks}

\author{\IEEEauthorblockN{\small Alireza Alizadeh, \textit{Student Member, IEEE}, Mai Vu, \textit{Senior Member, IEEE}}\\
%%\IEEEauthorblockA{Department of Electrical and Computer Engineering,
%%Tufts University, Medford, MA, USA\\
%%Email: \{alireza.alizadeh, mai.vu\}@tufts.edu}
%\and
%\IEEEauthorblockN{Mai Vu}
%\IEEEauthorblockA{Department of Electrical and Computer Engineering\\
%Tufts University\\
%Medford, USA\\
%Email: mai.vu@tufts.edu}
}


%Telephone: (800) 555--1212, Fax: (888) 555--1212}


\maketitle
\begin{abstract}
%\boldmath
User association is necessary in dense millimeter wave (mmWave) networks to determine which base station a user connects to in order to balance base station loads and maximize throughput. Given that mmWave connections are highly directional and vulnerable to small channel variations, user association changes these connections and hence significantly affects the user's instantaneous rate as well as network interference. In this paper, we introduce a new load balancing user association scheme for mmWave MIMO networks which considers this dependency on user association of user's transmission rates and network interference. We formulate the user association problem as mixed integer nonlinear programming and design a polynomial-time algorithm, called Worst Connection Swapping (WCS), to find a near-optimal solution. Simulation results confirm that the proposed user association scheme improves network performance significantly by moving the traffic of congested base stations to lightly-loaded ones and adjusting the interference accordingly. Further, the proposed WCS algorithm outperforms other generic algorithms for combinatorial programming such as the genetic algorithm in both accuracy and speed at several orders of magnitude faster, and for small networks where exhaustive search is possible it reaches the optimal solution.
\end{abstract}


% \fi
%\IEEEpeerreviewmaketitle

\section{Introduction}
%%%%%%%%%%%%%%%%%%%%%%%%%%%%%%%%%%%%%%%%%%%%%%
%%%%%%%%%%%%%%%%%%%%%%%%%%%%%%%%%%%%%%%%%%%%%%
%%%%%%%%%%%%%%%%%%%%%%%%%%%%%%%%%%%%%%%%%%%%%%
%\subsection{Motivation for mmWave}
\IEEEPARstart{B}{andwidth}
shortage is a major challenge for current wireless networks. Most of the available spectrum at microwave frequencies is occupied while there is a pressing need for higher throughputs and larger bandwidths \cite{Rapp_BW}. 
During the past few years, millimeter wave (mmWave) frequencies have attracted the interest of academia and industry due to the capability of multi-Gbps data rates and the huge amount of bandwidth available at frequencies between 30 - 300 GHz. 
MmWave communication for the new 5th generation (5G) cellular systems is a promising solution to the current spectrum shortage and ever-increasing capacity demand \cite{Cisco}, where its feasibility has been demonstrated not only for backhaul but also for mobile access \cite{RappWorks}-\cite{Roh}. 

User association plays an important role in resource allocation for cellular networks. 
%The goal of this procedure is to determine optimal connections between user equipments (UEs) and base stations (BSs) in order to achieve the best network performance. 
While user association has been studied in the context of heterogeneous networks (HetNets) \cite{Andrews}, massive multiple-input multiple-output (MIMO) systems \cite{Caire}, and mmWave Wi-Fi networks \cite{60GHz}, the problem in mmWave cellular networks is significantly different.
Because of the higher path loss and sensitivity to blockage, mmWave signals can only propagate a much shorter distance than the current RF signals, making it necessary for the mmWave cellular network to be dense. The base station (BS) density is expected to be higher by an order of magnitude \cite{UA_NYU}, and each user equipment (UE) is likely to be surrounded by a number of BSs, where the channel to each of these BS can vary significantly given the erratic nature of mmWave propagation with probabilistic line-of-sight (LOS), non-LOS (NLoS) and outage events. 
In addition, mmWave systems require the use of MIMO beamforming at both the transmitter and receiver ends \cite{Roh}, different from massive MIMO which has beamforming at only the BS. Hence the formulation of a user association problem for mmWave is different from other systems and requires a new approach.
%in that it must include joint transmit-receive beamforming. 

%While beamforming design (including both transmitter precoding and receiver combiner) is also a relevant problem related to user association, the joint optimization of both beamforming and association such as that studied in [9] may be too complex to be practical. Hence in this paper, we adopt the approach of assuming a given Tx-Rx beamforming design, which can be any combination of digital, analog or hybrid beams, and optimize for association.

\begin{comment}
\begin{figure}[]
\centering
\includegraphics[scale=.7]{UA1.png}
\caption{Load balancing user association. The dashed lines represent the previous connections that overload the macro BS, and the solid lines with arrowheads show the new connections after performing load balancing user association.}
\label{LBUA}
\end{figure}
\end{comment}

\subsection{Background and Related Work}
%need to make this part flow better, focus on what techniques have been proposed for what scenarios / systems and why they may not be applicable to mmWave networks.
%also in writing try not to say "the researchers did this or did that" -- focus on what was done not the people who did it.
\textit{Max-SINR} user association is a traditional method of using the highest signal to interference plus noise ratio (SINR) or largest received power to associate an UE with a BS \cite{3GPP2013}. This technique has been working well in cellular networks at microwave frequencies where all cells are homogeneous macro cells. The emergence of HetNets with smaller, lower power pico, femto and relay BSs requires a different look at user association to ensure load balancing across the network, avoiding the case that all users connect to only a macro BS because of its strongest SINR and overload it.

%There are different load balancing approaches based on moving the traffic to lightly-loaded BSs.
\textit{Cell breathing} is a technique that balances the load in a single tier network consisting of BSs of the same type by allowing each BS to adjust its transmit power, based on its load, to change its coverage area \cite{DynLB}-\cite{CellBreath}. If the BS is underloaded (overloaded), the transmit power increases (decreases) to expand (contract) the coverage region. HetNets also requires balancing the network load among different tiers. A conventional approach is \textit{biasing} in which the transmit power of BSs in a low-power lightly loaded tier is artificially biased in order to seem more attractive than congested BSs \cite{CellBias}. 
\textcolor{black}{
Although biasing can be done both within and among tiers, it cannot precisely control the load on each BS \cite{Biasing1}-\cite{Biasing2}.
}
%Biasing can only balance the traffic among the tiers, but not within each tier.
%, which makes it unsuitable for HetNets. Thus, we need a load balancing user association scheme that can balance the traffic among all the BS across the network. 

A load balancing user association scheme for HetNets uses an optimization theoretic approach to achieve the balance among all BSs across different tiers in the network \cite{Andrews}. This reference presented a novel user association technique with integer constraints on association coefficients and showed that this unique association problem is a combinatorial program, which is NP-hard. Considering a heuristic approach, the researchers first relaxed the unique association constraints to solve for a joint association problem (allowing a UE to be served by multiple BSs), and then rounded the relaxed solution. This relaxation reduced the complexity of the problem and provided an upper performance bound for the original unique association. %Finally, it it shown that the optimization problem is convex and can be solved via Lagrangian dual decomposition. 

%Optimizing user association is also studied in a 60-GHz mmWave Wi-Fi network \cite{60GHz}, at which high atmospheric-absorption frequency, the user instantaneous rates converge to deterministic values and the interference is assumed to be negligible due to directional steerable antenna arrays. 
%This work is one of the first works which studied a fundamental 

In a similar work to \cite{Andrews}, the problem of user association in HetNets with massive MIMO BSs and single-antenna UEs is studied in \cite{Caire}. Because of the channel hardening effect in massive MIMO, the considered association approach is independent of the user instantaneous rate or instantaneous channel state information (CSI), but only depends on converging deterministic rate values as a function of the system parameters and large-scale CSI. 
Even though unique association is considered, the association coefficients are not necessarily integers, but represent the limit of the fraction of time slots in which each UE is served by one BS. 
Using this interpretation of fractional user association as long-term time average, the researchers then formulated user association as a convex network utility maximization problem and solved it using Lagrangian duality. 

Optimizing user association is also studied in a 60-GHz mmWave Wi-Fi network \cite{60GHz}. The optimization problem in \cite{60GHz} is similar to the one appeared in \cite{Andrews}, but instead of maximizing network throughput it is formulated to minimize the maximum per-BS load subject to demanded user data rates.
At this high atmospheric-absorption frequency, the user instantaneous rates converge to deterministic values and the interference is assumed to be negligible due to directional steerable antenna arrays. This assumption simplifies the user association problem compared to the case when we must take into account the effect of interference, as in a dense mmWave cellular network at the proposed frequencies (28, 38, and 73 GHz), for which interference has been noted for outdoor environments \cite{RappWorks}-\cite{Akendiz}, \cite{Niu}.

A number of other existing works also considered the joint problem of user association, beamforming design, and power allocation \cite{Hong}-\cite{Razaviyan}. This joint problem is shown to be NP-hard, and researchers usually proposed iterative algorithms to achieve near-optimal solutions. Algorithms solving such an NP-hard joint optimization problem may be too complex to be implemented in practice. As such, we take the alternative approach of fixing the beamforming designs and optimize for user association alone. 
%None of the approaches in the previous works is applicable to mmWave systems in the sense of following aspects:

\textcolor{black}{
Most of recent works on user association ignore the effect of small-scale fading (instantaneous CSI) on the user instantaneous rate and consider a scalar channel model with only large-scale fading \cite{R3_1}-\cite{R2_4}, hence the SINR used for user association is only a function of large-scale parameters (e.g. distance) regardless of small-scale channel variations. Specifically, the SINR term in the capacity formula is not a matrix but a scalar value indicating that the mmWave MIMO small-scale variations were ignored and  the system is usually simplified by only considering directional gains for mmWave antenna. This consideration reduces the complexity of the analysis, but this approach is not suitable for mmWave systems where the channel variation can be fast and the instantaneous CSI may change rapidly \cite{UA_NYU}. Our simulation results confirm that considering both large-scale and small-scale CSI significantly improves the average network spectral efficiently.}

%This \textcolor{blue}{consideration} \textcolor{red}{\st{fixed user rate approach}} also led to a simpler interference structure, in which the interference from other BSs (in the downlink for example) is assumed to be both independent of user association and present all the time (full interference). This assumption is also not realistic in mmWave networks, where strong directionality of beamforming transmission and reception makes the interference highly dependent on which UEs connect to which BSs and may not always be fully present at all time. 

%The problem of unique association is also relaxed due to its combinatorial structure. These works also solved the joint association problem which may impose significant hardware overhead, and hence may not be practical to implement.

\subsection{Our contributions}
In this paper we introduce a new user association problem formulation suitable for mmWave networks. Specifically, unlike the typical approach in user association for HetNets \cite{Andrews}, massive MIMO \cite{Caire}, and mmWave Wi-Fi \cite{60GHz} which accounts only for large-scale CSI, we consider both the large-scale and instantaneous CSI in our formulations. This consideration is suitable for mmWave networks where the channels can vary significantly even during two consecutive time slots.
\textcolor{black}{
In the case that instantaneous CSI is unavailable, however, we can still use the proposed user association scheme but only with large-scale CSI which will provide a much longer time span. 
Adaptation to instantaneous CSI will further improve the network performance. As our user association scheme is independent of channel model, the ideal assumption of having both large-scale and instantaneous CSI can be considered as a performance upper bound. In practical scenarios we may use the proposed user association scheme with only large-scale CSI or with both types of CSI depending on the availability.
}

Furthermore, previous works ignore the effect of user association on network interference; consequently in the downlink for example, each user suffers full interference from all other BSs, regardless of
their association. While this assumption makes sense in an omnidirectional transmission, it no
longer holds in a directional transmission as is the case in mmWave, where the interference
will change substantially depending on which UE beamforms with which BS. The network interference structure in mmWave is highly dependent on user association and we need to consider this dependency while computing the user rates. Not adapting the interference to user association unfortunately degrades the network spectral efficiency severely. 

The main contributions of this paper are summarized as follows: 
\begin{itemize}
\item We propose a new load balancing user association scheme which takes into account the dependency of network interference on user association. 
In particular, we formulate the user instantaneous rate as a function of user association, as is the case in mmWave.
Consequently, the total interference coming from other BSs (while serving other UEs) also depends
on association coefficients. 
\item We introduce an \textit{Activation Matrix} which specifies the BS each UE connects to during each time slot, from which association coefficients are derived as its time average. 
The activation matrix allows per-time-slot user association which is necessary because the instantaneous CSI of a mmWave channel may change significantly even during two consecutive time slots.
\item For the proposed user association scheme, we formulate an optimization problem as a mixed integer nonlinear programming (MINLP), which is known to be NP-hard due to its non-convex nonlinear structure and presence of integer variables.  
This optimization problem is similar in form to those derived in \cite{Andrews}-\cite{Caire}, but instead of being a linear function of the user association coefficients, its objective function is highly dependent on the network interference structure and user association in a non-linear fashion. This makes the optimization problem highly non-convex and challenging, which in turn makes our proposed user association scheme suitable for mmWave cellular networks and reveals its novelty.
\item We design a polynomial-time specialized efficient algorithm, called Worst Connection Swapping (WCS), to solve the formulated MINLP and find a near-optimal solution for the user association. 
This algorithm is based on the intuition that swapping the worst UE-BS connection is likely to provide the UE a stronger link to another BS and/or reduce the interference, which consequently improves the user's rate.
Complexity analysis reveals that our proposed algorithm's runtime grows as $K^2 \log(K) M^2$ with $K$ as the number of UEs and $M$ as the number of antennas at each BS.
\item Our simulation results show that considering instantaneous CSI significantly improves the network spectral efficiency. Further, the proposed load balancing scheme outperforms max-SINR and other load balancing schemes by including the dependency of interference on user association. Numerical results also confirm the analysis of algorithm complexity and show that the WCS algorithm has fast convergence with the number of iterations growing linearly with the number of UEs, and the runtime growing quadratically, which is several orders of magnitude faster than genetic algorithm (GA) and exhaustive search.
\textcolor{black}{
\item 
Our proposed user association scheme can be integrated with any mobility model since it is designed to be performed per-time slot. As an illustration, we provide simulation results of our scheme applied to a simple user mobility model where the UEs have fixed locations during each time slot, and in the consecutive time slot each UE moves to a new random location within a given range from its location in the previous time slot. Our simulations show the effect of mobility on the association results, for example, a UE with strong signal from a particular BS tends to be associated with that BS even as it moves around within a vicinity, whereas a UE with comparable signals from several BSs can switch association frequently as it mobilizes.
}
\end{itemize}

\subsection{Paper Organization}
The rest of the paper is organized as follows. We start with the channel and system model in Section II. In Section III, we introduce our load balancing user association scheme and define the activation and association matrices. The user association optimization problem is formulated in Section IV. In Section V, we introduce the WCS algorithm and analyze its complexity. We present the numerical results in Section VI, and Section VII contains our conclusion.

\subsection{Notation}
Throughout this paper, scalars are represented by lowercase letters, vectors are denoted by lowercase boldface letters and matrices by uppercase boldface letters. Superscript $(.)^*$ denotes the conjugate transpose, $\log(.)$ stands for base-2 logarithm, $\text{mod}(.)$ represents the modulo operation, and big-O notation $\mathcal{O}(.)$ expresses the complexity.
 $\boldsymbol{I}_N$ is the $N\times N$ identity matrix, and we use $\mathrm{tr}(\mathbf{A})$ and $|\mathbf{A}|$ to denote the trace and determinant of matrix $\mathbf{A}$, respectively. The distribution of a complex Gaussian vector $\mathbf{x}$ with mean $\mathbold{\mu}$ and covariance matrix $\mathbf{Q}$ is denoted by $\mathbf{x}\sim\mathcal{CN}(\mathbold{\mu},\mathbf{Q})$.

%and we use $\mathrm{tr}(\boldsymbol{A})$ and $\det(\boldsymbol{A})$ to denote the trace and the determinant of matrix $\boldsymbol{A}$, respectively. 
%The operator $\mathcal{E}[.]$ stands for the expectation
%and $z\sim\mathcal{CN}(0,1)$ is a circularly symmetric complex Gaussian random variable, where its real and imaginary parts are zero mean i.i.d. Gaussian random variables, each with variance 1/2, i.e., $\mathcal{N}(0,1/2)$.

\section{Channel and System Model}
\subsection{mmWave Channel Model}
The mmWave channel has completely different characteristics compared to the typical microwave model of a rich scattering independent and identically distributed (i.i.d.) channel. 
The channel model considered in this paper is based on the clustered channel model introduced in \cite{SS} and the 3GPP-style 3D channel model proposed for the urban micro (UMi) environments in \cite{Nokia}, which is based on the measurements obtained in New York City.
This channel model has $C$ clusters with $L$ rays per cluster, and it can be expressed as
\begin{align}\label{clustered_ch}
H=\frac{1}{\sqrt{CL}}\sum_{c=1}^{C}\sum_{l=1}^{L} \sqrt{\gamma_c}~\mathbf{a}(\phi_{c,l}^{\textrm{UE}},\theta_{c,l}^{\textrm{UE}}) ~\mathbf{a}^*(\phi_{c,l}^{\textrm{BS}},\theta_{c,l}^{\textrm{BS}})
\end{align}
where $\gamma_c$ is the power gain of the $c$th cluster. The parameters $\phi^{\textrm{UE}}$, $\theta^\textrm{UE}$, $\phi^\textrm{BS}$, $\theta^\textrm{BS}$ represent azimuth angle of arrival (AoA), elevation angle of arrival (EoA), azimuth angle of departure (AoD), and elevation angle of departure (EoD), respectively. These parameters are generated randomly based on different distributions and cross correlations as given in \cite[Tables 1-3]{Nokia}. The vector $\mathbf{a}(\phi,\theta)$ is the antenna array response vector, which depends on antenna geometry, including uniform linear array (ULA) or uniform planar array (UPA). In order to enable beamforming in the elevation direction (3D beamforming), we use the uniform $U\times V$ planar array given by \cite{SS}
%\frac{1}{\sqrt{S}}
\begin{align}
\mathbf{a}(\phi,\theta)=&\big[ 1, ..., e^{jkd_{\textrm{a}}(u\sin(\phi)\sin(\theta)+v\cos(\theta))}, ...,\nonumber \\  &e^{jkd_{\textrm{a}}((U-1)\sin(\phi)\sin(\theta)+(V-1)\cos(\theta))} \big]^T
\end{align}
where $d_a$ is the distance between antenna elements, and $u\in\{1, ..., U\}$ and $v\in\{1, ..., V\}$ are the indices of antenna elements.

%, and $S=UV$ is the antenna array size.

%Some of the channel parameters are delay spread, azimuth angle of departure (AoD) spread, azimuth angle of arrival (AoA) spread, elevation angle of departure (EoD) spread, elevation angle of arrival (EoA) spread, EoD bias, EoA bias, line of sight (LoS) shadow fading, (non-line of sight) NLoS shadow fading. These parameters are generated 
We consider two link states for each channel, LoS and NLoS, and use the following probability functions obtained based on the New York City measurements in \cite{RapLetter}
\begin{align}\label{p_LoS/NLoS}
p_{\text{LoS}}(d)&=\Big[\min\Big(\frac{d_\textrm{BP}}{d},1\Big).\Big(1-e^{-\frac{d}{\eta}}\Big)+e^{-\frac{d}{\eta}} \Big]^2\\
p_{\text{NLoS}}(d)&=1-p_{\text{LoS}}(d)
\end{align}
where $d$ is the 3D distance between UE and BS in meters, $d_{\textrm{BP}}$ is the breakpoint distance at which the LoS probability is no longer equal to 1, and $\eta$ is a decay parameter. The obtained values based on measurements for these parameters are $d_{\textrm{BP}}=27$ m and $\eta=71$ m.

%The probability model in () is obtained based on the New York City measurements collected in [3,4 of Rappaport, letter].
Moreover, we use the following path loss model for LoS and NLoS links \cite{Nokia}
 % Sigma_SF = 4.9*randn Rappaport's paper Eq. (1) LoS
\begin{align}\label{PL}
PL[\textrm{dB}]=20\log_{10}\Big(\frac{4\pi d_0}{ \lambda}\Big) + 10n \log_{10}\Big(\frac{d}{d_0}\Big) + X_{\sigma_{\textrm{SF}}}
\end{align}
where $\lambda$ is the wavelength, $d_0$ is the reference distance, $n$ is the path loss exponent, and 
$X_{\sigma_{\textrm{SF}}}$ is the lognormal random variable with standard deviation $\sigma_{\textrm{SF}}$ (dB) which describes the shadow fading.
%\begin{align}
%PL(dB)=20\log_{10}\Big(\frac{4\pi d_0}{ \lambda}\Big) + 10n \log_{10}\Big(\frac{d}{d_0}\Big) + \sigma_{\textrm{SF}}
%\end{align}
These parameters vary depending on LoS of NLoS propagation at 73 GHz, the path loss exponents and the shadowing factors are $n_{\textrm{LoS}}=2$, $n_{\textrm{NLoS}}=3.4$, $\sigma_{\textrm{SF, LoS}}=4.8$ dB, and $\sigma_{\textrm{SF, NLoS}}=7.9$ dB.

In 4G cellular networks, pilot signals are used to estimate CSI at the receiver. Once the CSI is available at the receiver, it can be shared with the transmitter via limited feedback or channel reciprocity. 
However, in 5G dense HetNets these conventional approaches are inapplicable due to network densification and the limited amount of pilot resources \cite{User-centric C-RAN}. To address the new challenges emerging from network densification, a promising radio access technology, called cloud radio access network (C-RAN), has been proposed \cite{HetNet C-RAN}. In this radio network, CSI at both the transmitter and the receiver can be estimated through new CSI acquisition schemes and shared via C-RAN for centralized signal processing, coordinated beamforming, and resource allocation in 5G new radio \cite{Large-Scale C-RAN}, \cite{5GNR}. In this paper, we assume that the CSI is estimated for the aforementioned applications and we can utilize it for the purpose of user association. 
\textcolor{black}{
We note that C-RAN is an architecture that enables this shared CSI and centralized user association, but it may not be the only one, and the CSI required for beamforming in 5G can be used for user association. Further, distributed algorithms can also be developed for the same problem formulation which we will consider here, but do not require C-RAN because of the distributed nature.
}

Different user association schemes make use of different types of channel models: when an algorithm uses the detailed clustered channel model in (\ref{clustered_ch}), we call this \emph{instantaneous CSI}, and when an algorithm uses only path loss, shadowing and LOS/NLOS channel models as in (\ref{p_LoS/NLoS})-(\ref{PL}), we call this \emph{large-scale CSI}. 
\textcolor{black}{
Our proposed user association scheme as discussed in Section III utilizes both large-scale and instantaneous CSI as given in (\ref{clustered_ch}) . The scheme, however, is also applicable if the channel model instead contains only large-scale fading as considered in [8, eq. (1)].
%For this case, we generate only one instantaneous CSI and keep it fixed for all UE-BS channels, and then multiply it with large-scale fading including path loss, shadowing and LOS/NLOS information as in (\ref{p_LoS/NLoS})-(\ref{PL})).
%*** add here discussion on how our scheme only uses LSF -- how is the channel matrix modeled in that case? ***
In the simulation section, we show examples of both cases and provide comparison with other schemes which mostly rely on large-scale CSI.
}

%is the small-scale fading and βjlk is the large-scale fading coefficient. Here, we make the block fading assumption that the large-scale fading coefficients are kept fixed over lots of coherence time intervals and also assume that large-scale fading coefficients are known at the BS [5], while small-scale fading fading coefficients remain fixed within a coherence time interval. At the same time, each user’s channel is considered to be independent from other users’ channels.

%We assume each time slot to be short enough such that the small-scale fading of the channel remains constant within a time slot, while 
\subsection{System and Signal Model}
%%%%%%%%%%%%%%%%%%%%%%%%%%%%%%%%%%%%%%%%%%%%%%
%%%%%%%%%%%%%%%%%%%%%%%%%%%%%%%%%%%%%%%%%%%%%%
%%%%%%%%%%%%%%%%%%%%%%%%%%%%%%%%%%%%%%%%%%%%%%
We consider a downlink scenario in a cellular mmWave MIMO network with $J$ BSs and $K$ UEs. $M_j$ and $N_k$ are the number of antennas at BS $j$ and UE $k$, respectively, where we assume $M_j \geq N_k$, as handsets usually have fewer antennas than BSs. Let $\mathcal{J}=\{1, ..., J\}$ denotes the set of BSs and $\mathcal{K}=\{1, ..., K\}$ represents the set of UEs. 
Each UE $k$ aims to receive $n_k$ data streams from its serving BS such that $1\leq n_k\leq N_k$, where the upper inequality comes from the fact that the number of data streams for each UE cannot exceed the number of its  antennas. 

Thus, we can define the total number of downlink data streams sent by BS $j$ as
%on any given time slot as
\begin{equation}
D_j=\sum_{k \in \mathcal{Q}_j(t)}n_k
\end{equation}
We define $\mathcal{Q}_j(t)$ as the \textit{Activation Set} which represents the set of active UEs in BS $j$ within time slot $t$, such that  $\mathcal{Q}_j(t) \subseteq \mathcal{K}$ and $|\mathcal{Q}_j(t)|=Q_j(t)\leq K$. Note that the total number of downlink data streams sent by each BS should be less than or equal to its number of antennas, i.e., $D_j \leq M_j$. For notational simplicity, we drop the time index $t$ in definition of $D_j$, and only keep the time index for $Q_j(t)$ due to its importance.
%Note that for single antenna UEs ($n_k=N_k=1,~\forall k \in \mathcal{K}$), the total number of downlink data streams of each BS is equal to the number of its active UE, i.e., $D_j= Q_j(t)$.

%%%%%%%%%%%%%%%%%%%%%%%%%%%%%%%%%%%%%%%%%%%%%%%%%
The $M_j\times 1$ transmitted signal from BS $j$ is given by
\begin{equation}\label{x_j}
\mathbf{x}_j = \mathbf{F}_j \mathbf{d}_j = \sum_{k\in \mathcal{Q}_j(t)}\mathbf{F}_{k,j}\mathbf{s}_k
\end{equation}
where $\mathbf{s}_k\in \mathbb{C}^{n_k}$ is the data stream vector for UE $k$ consists of mutually uncorrelated zero-mean symbols, with $\mathbb{E}\lbrack \mathbf{s}_k\mathbf{s}_k^*\rbrack = \mathbf{I}_{n_k}
$. The column vector $\mathbf{d}_j\in \mathbb{C}^{D_j}$ represents the vector of data symbols of BS $j$, which is the vertical concatenation of the data stream vectors $\mathbf{s}_k,~k\in\mathcal{Q}_j(t)$, such that $\mathbb{E}\lbrack \mathbf{d}_j\mathbf{d}_j^*\rbrack = \mathbf{I}_{D_j}$.
Matrix $\mathbf{F}_{k,j}\in\mathbb{C}^{M_j\times n_k}$ is the linear precoder for each UE $k$ associated with BS $j$, and $\mathbf{F}_j\in\mathbb{C}^{M_j \times D_j}$ is the complete linear precoder matrix of BS $j$ which is the horizontal concatenation of all $\mathbf{F}_{k,j}, k\in\mathcal{Q}_j(t)$.

The power constraint at BS $j$ can be described as 
\begin{equation}\label{power}
\mathbb{E}[\mathbf{x}_j^* \mathbf{x}_j]=\sum_{k\in \mathcal{Q}_j(t)}\textrm{Tr}(\mathbf{F}_{k,j}\mathbf{F}_{k,j}^*)\leq P_j 
\end{equation}
where $P_j$ is the transmit power of BS $j$.
Now we can express the $N_k\times 1$ received signal at UE $k$ antennas as
\begin{equation}\label{y_k}
\mathbf{y}_k = \sum_{j\in \mathcal{J}}\mathbf{H}_{k,j}\mathbf{x}_j + \mathbf{z}_k,
\end{equation}
and the post-processed signal received by each UE is
\begin{equation}\label{y_tilde_k}
\tilde{\mathbf{y}}_k = \sum_{j\in \mathcal{J}}\mathbf{W}_k^* \mathbf{H}_{k,j}\mathbf{x}_j + \mathbf{W}_k^*\mathbf{z}_k
\end{equation}
%(including both small- and large-scale fading components)
where $\mathbf{W}_k\in\mathbb{C}^{N_k \times n_k}$ is the linear combiner matrix of UE $k$, $\mathbf{H}_{k,j}\in\mathbb{C}^{N_k\times M_j}$ represents the channel matrix between BS $j$ and UE $k$, and $\mathbf{z}_k\in\mathbb{C}^{N_k}$ is the white Gaussian noise vector at UE $k$, with $\mathbf{z}_k\sim\mathcal{CN}(\mathbf{0},N_0 \mathbf{I}_{N_k})$.
%It is worth mentioning that in MIMO mmWave systems hybrid (analog and digital) beamforming should be implemented to reduces cost and power consumption of large antenna arrays \cite{SS}.
The presented signal models (\ref{x_j})-(\ref{y_tilde_k}) are applicable for all types of transmit beamforming and receive combining. Later for user association optimization, we will specify the transmit and receive beamforming schemes. Further, based on C-RAN architecture as discussed earlier, each BS knows its channels to all UEs and can share that CSI with all other BSs for both beamforming design and user association purposes.
\section{MmWave Load Balancing User Association}
%Association and Activation Matrices
In the literature, when computing the instantaneous rate for a specific UE (connected to a BS), the interference coming from other UE-BS connections is assumed to be both independent of the user association and present all the time (full interference) \cite{Andrews}-\cite{60GHz}. This assumption is not realistic in mmWave systems due to the use of beamforming and results in lower instantaneous user rates. In mmWave systems, the network interference highly depends on user association and we need to consider this while computing user rates. Moreover, user association depends on channel realizations which is highly directional and can vary fast in mmWave frequencies. Thus, we cannot use the full interference structure in mmWave systems. 

In this section, we introduce a new user association model for mmWave MIMO networks. First, we need to introduce our definition of a \textit{time slot} throughout this paper. Each time slot $t$ is a duration of time comparable to channel coherence time such that the instantaneous CSI remains relatively the same within it and only changes from one time slot to another.
During time slot $t$, each UE is connected to exactly one BS. Because of beamforming, the interference structure in each time slot depends on the user association on that specific slot. This interference structure is appropriate for mmWave channels where the channel is probabilistic and can vary fast.
Moreover, we assume it is possible to split the data streams of each UE and transmit them in different time slots. 
Considering above definitions, we study two association approaches in this paper: (i) instantaneous user association, which is performed within each time slot and results in unique association (each UE can be associated with only one BS during each time slot), and (ii) fractional (joint) user association, which is obtained by averaging the instantaneous association over $T$ time slots (for an arbitrarily large $T$). At each time slot, the mmWave channels are generated independently based on the channel model presented in Section II-A. 
%Note that performing instantaneous user association may not be practical in each time slot due to the very short channel coherence time in mmWave systems, but 
We consider both approaches to evaluate the performance of our proposed user association model and compare it with existing user association schemes.
%In section V, we evaluate the performance of the TFA method in both cases.

\textcolor{black}{
It is worth mentioning that our proposed per-time-slot user association can be carried out under the synchronizations requirement of modern cellular systems. In LTE TDD networks, the basic synchronization requirement for any pair of cells with overlapping coverage areas which operate on the same frequency is $3~\mu s$ \cite{3GPP2017}, and it would be in the order of $1 ~\mu s$ or below for 5G systems \cite{3GPP2018}. Moreover, based on the work in \cite{5GmmWaveNYU2016}, a time slot of duration $T_{slot}=100~\mu s$ is considered as a sufficiently short time duration to ensure channel coherence at mmWave systems. Since the synchronization time is relatively small compared to the time slot duration, synchronization will not be an issue for performing per-time-slot user association.
}

\textcolor{black}{
We differentiate between user association and user scheduling. Note that user association determines the UE-BS connections among multiple BSs, while user scheduling allocates the available resources (in this case, data streams) at each BS to its associated UEs. Thus, user scheduling can be considered as a post-association stage.
In the literature, it is shown that the joint problem of user association and user scheduling has high computational complexity and researchers usually decouple the joint optimization problem into sub-problems to find the optimal solutions \cite{Andrews},  \cite{Joint_US_UA2014}. 
In this paper, we assume that the BSs' loads are determined based on their available resources. Specifically, we consider the multi-user MIMO scenario where at each time slot each BS can simultaneously transmit to multiple UEs by constraining the available number of data streams at each BS to be smaller than the number of its antennas.
Therefore, at each time slot, the BS can serve all its associated users simultaneously, by beamforming simultaneously to multiple users using multiple beams, and there are enough spatial degrees of freedom for all the associated users as long as the aforementioned constraint is satisfied. As such, in this case, there is no need to perform scheduling among the associated users. 
In the case that the number of required data streams at each time slot is larger than the number of BS antennas, then user scheduling is inevitable. In such a case, a simple scheduling policy (e.g. round robin scheduling \cite{Joint_US_UA2014}) or more efficient scheduling schemes (e.g. proportional fairness scheduling \cite{Andrews} or CDF-based scheduling \cite{Scheduling2005}, \cite{Scheduling2016}) can be implemented to allocate resources of each BS among its associated users.
}

%At this point, we need to introduce our definition of \textit{time block} and \textit{time slot} throughout this paper. Each time block of length $T$ consists of many time slots $t$ which are considered to be short enough such that the small-scale fading characteristics of the channel remains constant within a time slot. Moreover, we assume the large-scale fading characteristics (path loss and shadowing) of the channel are kept fixed over each time block and they only change from one time block to another. Considering above definitions, we study two association approaches in this paper: 1) Instantaneous user association which is performed within each time slot and results in unique association, and 2) long-term user association which is done over a time block and leads to joint user association. In section V, we evaluate the performance of the TFA method in both cases.

We start by defining the \textit{Activation Matrix} as
\begin{equation}
\mathbf{B}\triangleq
\left[ \begin{array}{ccc}
\mathbold{\beta}(1)&\cdots &\mathbold{\beta}(T) 
\end{array}
\right]=
\left[ \begin{array}{ccc}
\beta_1(1) & \cdots & \beta_1(T) \\
\vdots & \ddots &\vdots \\
\beta_K(1) & \cdots & \beta_K(T) \\
\end{array} \right]
\end{equation}
where $\mathbold{\beta}(t)$ is called the \textit{Activation Vector} at time slot $t$, and each element of $\mathbf{B}$ is called an \emph{activation factor} and is the index of BS to whom user $k$ is associated with during time slot $t$, i.e., $\beta_k(t)\in\mathcal{J}$ with $k\in\mathcal{K}$ and $t\in\mathcal{T}=\{1, ..., T\}$.  
Considering above definition, the relationship between the activation set of BS $j$ and the activation factors can be described as
\begin{equation}\label{Q_j}
\mathcal{Q}_j(t) = \{ k: \beta_k(t)=j\}.
\end{equation}
As stated earlier, we assume each UE can be associated with only one BS at any time slot $t$, i.e.,
\begin{equation}
\mathcal{Q}_j(t) \cap \mathcal{Q}_i(t) = \varnothing,~~j\neq i
\end{equation}
\begin{equation}\label{union}
\bigcup\limits_{j=1}^{J}\mathcal{Q}_j(t) = \mathcal{K}
\end{equation}
where (\ref{union}) indicates that during each time slot all UEs are served by the BSs.

The elements of activation matrix should satisfy the following conditions
\begin{align}
\sum_{j\in\mathcal{J}} 1_{\beta_k(t)}(j) &\leq 1, ~~\forall k\in \mathcal{K}\label{TFA_cons_1}\\
\sum_{k\in\mathcal{K}}1_{\beta_k(t)}(j). n_k &\leq D_j, ~~\forall j\in \mathcal{J}\label{TFA_cons_2}
\end{align}
where the indicator function is defined as 
\begin{align}
1_{\beta_k(t)}(j)\triangleq
\begin{cases}
   1,& ~~~~ \beta_k(t)=j\\
   0,& ~~~~ \beta_k(t)\neq j
\end{cases}
\end{align}
The activation constraints in (\ref{TFA_cons_1}) reflect the fact that each UE cannot be associated with more than one BS in each time slot, and the resource allocation constraints in (\ref{TFA_cons_2}) denote that the sum of data streams of UEs served by each BS cannot exceed the total number of available data streams on that BS. Note that $1_{\beta_k(t)}(j)$ is equal to one only if $\beta_k(t)=j$ or equivalently, $k\in\mathcal{Q}_j(t)$. Thus, the summation in (\ref{TFA_cons_2}) is actually over the set of active UEs in BS $j$.

Next, we define the \textit{Association Matrix} $\mathbf{A}$ as follows
\begin{equation}
\mathbf{A}\triangleq
\left[ \begin{array}{ccc}
\alpha_{1,1} & \cdots & \alpha_{1,J} \\
\vdots & \ddots &\vdots \\
\alpha_{K,1} & \cdots & \alpha_{K,J} \\
\end{array} \right]
\end{equation}
where $\alpha_{k,j}\in [0,1]$ is the (fractional) association coefficient representing the average connectivity of UE $k$ to BS $j$. If $\alpha_{k,j}=0$, we say UE $k$ is not associated with BS $j$ over time. 
In this model, association coefficients are considered as a fraction of time. Specifically, $\alpha_{k,j}$ is the average fraction of time during which UE $k$ is connected to BS $j$. 
The relationship between association coefficients and activation factors is given by
\begin{equation}\label{alpha_beta}
\alpha_{k,j} = \lim_{T\rightarrow \infty} \frac{1}{T}\sum_{t=1}^T 1_{\beta_k(t)}(j).
\end{equation}
According to (\ref{alpha_beta}), given the activation matrix $\mathbf{B}$, one can easily obtain the association matrix $\mathbf{A}$. For this reason, we formulate our user association problem in terms of the activation factors, which are integer variables representing the serving BS indices.

%%%%%% MOBILITY & HANDOVER%%%%%%
\textcolor{black}{
Since the proposed user association scheme is carried out per-time-slot, it can result in frequent handover, especially in the case that the UE is mobile and has comparable signal strengths from several surrounding BSs. While we do not consider handover cost explicitly in this work, we note that the proposed user association can be applied using only large-scale CSI which will result in a longer time span for each association and lessen handover. 
In Section VI, we consider a simple mobility model where the location of each UE is fixed during each time slot, and changes in the subsequent time slot to a new random location within a fixed range from its previous location. The user association is then performed at each time slot to assess the effect of mobility on association and handover.
}

\section{User Association Optimization Problem}
In this section, we first evaluate the instantaneous and average per-user throughputs by processing the received signal at each user. Then, we formulate an optimization problem to find the optimal activation factors that maximize the throughputs while satisfying load balancing constraints on the BSs.
\subsection{Formulation of instantaneous user rate}
Considering (\ref{x_j}), the received signal by UE $k$ in  (\ref{y_tilde_k}) can be decomposed as
\begin{align}\label{y_tilde_k_t}
\tilde{\mathbf{y}}_k (t)&= \underbrace{\mathbf{W}_k^*\mathbf{H}_{k,j}\mathbf{F}_{k,j} \mathbf{s}_k}_\text{Desired signal} + \underbrace{\mathbf{W}_k^*\mathbf{H}_{k,j}\sum_{\substack{l\in \mathcal{Q}_j(t) \\ l\neq k}}\mathbf{F}_{l,j} \mathbf{s}_l}_\text{Intra-cell interference} \nonumber \\
&+ \underbrace{\mathbf{W}_k^*\sum_{\substack{i\in \mathcal{J} \\ i\neq j}}\sum_{\substack{l\in \mathcal{Q}_i(t)}} \mathbf{H}_{k,i} \mathbf{F}_{l,i} \mathbf{s}_l}_\text{Inter-cell interference} + \underbrace{\mathbf{W}_k^*\mathbf{z}_k}_\text{Noise}
\end{align}
% \sum_{j\in \mathcal{J}}\mathbf{W}_k^* \mathbf{H}_{k,j} \mathbf{x}_j + \mathbf{W}_k^* \mathbf{z}_k\nonumber \\
where the first term is the desired received signal from the serving BS $j$, the second term represents the interference coming from the same BS $j$ by signals intended for its other active UEs, the third term is the interference coming from other BSs $i\neq j$ by signals sent to their active UEs, and the last term is the thermal noise at UE $k$.
The activation sets $\mathcal{Q}_j(t)$ and $\mathcal{Q}_i(t)$ appeared in the interference terms indicate that the interference is highly dependent on the user association, which highlights the novelty of this work.
Again, we note that all vectors and matrices in (\ref{y_tilde_k_t}) are time-dependent, and the time index $t$ is dropped for the sake of notational simplicity. 

When UE $k$ is connected to BS $j$ in time slot $t$, its instantaneous rate can be obtained as \cite{Telatar}
\begin{equation}\label{R_kj}
R_{k,j}(t) = \log_2\Big |\mathbf{I}_{n_k} + (\mathbf{V}_{k,j}(t))^{-1}\mathbf{W}_{k}^*\mathbf{H}_{k,j}\mathbf{F}_{k,j} \mathbf{F}_{k,j}^* \mathbf{H}_{k,j}^*\mathbf{W}_{k}\Big |
\end{equation}
where $\mathbf{V}_{k,j}$ is the interference-plus-noise covariance matrix given as
\begin{align}\label{Y_interf}
&\mathbf{V}_{k,j}(t)= \mathbf{W}_{k}^*\mathbf{H}_{k,j}\Big( \sum_{\substack{l\in \mathcal{Q}_{j}(t) \\ l\neq k}} \mathbf{F}_{l,j} \mathbf{F}_{l,j}^* \Big ) \mathbf{H}_{k,j}^*\mathbf{W}_{k} \nonumber \\
&+ \mathbf{W}_{k}^* \Big( \sum_{\substack{i\in \mathcal{J} \\ i\neq j}} \sum_{\substack{l\in \mathcal{Q}_i(t)}} \mathbf{H}_{k,i}\mathbf{F}_{l,i} \mathbf{F}_{l,i}^* \mathbf{H}_{k,i}^* \Big ) \mathbf{W}_{k} + N_0 \mathbf{W}_k^*\mathbf{W}_k.
\end{align}
% \nonumber \\ &
\begin{comment}
Considering (\ref{B}) and (\ref{X}), we can rewrite the instantaneous rate in compact form as
\begin{equation}\label{inst_rate_t}
R_{k,j}(t) = \log_2\det (\mathbf{I}_{N_k} + (\mathbf{Y}_{k,j}(t))^{-1}\mathbf{G}_{k,j})
\end{equation}
\begin{align}
\mathbf{Y}_{k,j}(t) &= \sum_{\substack{l\in \mathcal{Q}_{j}(t) \\ l\neq k}} \mathbf{X}_{l,j,k} + \sum_{\substack{i\in \mathcal{J} \\ i\neq j}} \sum_{\substack{l\in \mathcal{Q}_i(t)}} \mathbf{X}_{l,i,k} + N_0 \mathbf{W}_k^*\mathbf{W}_k
\end{align}
\end{comment}
The instantaneous rate given in (\ref{R_kj}) is a function of activation sets $\mathcal{Q}_j(t)$. Thus, the instantaneous per-user throughput at time slot $t$ can be expressed as 
\begin{align}\label{r_k_t}
r_{k}(t)=\sum_{j\in\mathcal{J}} 1_{\mathbold{\beta}_k(t)}(j)\times R_{k,j}(t),
\end{align}
and the average per-user throughput is given by
\begin{align}
r_{k}=\lim_{T\rightarrow \infty}\frac{1}{T}\sum_{t=1}^{T} r_k(t).
\end{align}
%\frac{1}{T}\sum_{j=1}^{J} \sum_{t=1}^{T} 1_{\mathcal{Q}_j(t)}(k)\times R_{k,j}(t)\\
\subsection{Optimization Problem}
As stated before, the channel variation can be fast and unpredictable in mmWave frequencies and the clustered channel as modeled in (\ref{clustered_ch}) may change significantly even during two consecutive time slots. Thus, we need to perform the user association in each time slot, which is comparable with the channel coherence time. 
Defining the instantaneous user throughput vector $\mathbf{r}(t)\triangleq (r_1(t), ..., r_K(t))$, we wish to find the optimal activation vector $\mathbold{\beta}(t)$ which maximizes an overall network utility function. This utility function should be concave and monotonically increasing. 
\textcolor{black}{
Our proposed user association scheme can implicitly incorporate user fairness via the choice of network utility function including the family of utility functions defined in \cite{Caire}.
}

\textcolor{black}{
In this paper, we consider two well-known and widely used utility functions. The \textit{sum-rate} utility function defined as
%  defined in (\ref{u(r(t))})
\begin{align}\label{u(r(t))}
U_s(\mathbf{r}(t))&\triangleq \sum_{k\in\mathcal{K}} r_k(t)
\end{align}
and the \textit{min-rate} utility function given by
\begin{align}\label{u2(r(t))}
U_m(\mathbf{r}(t))&\triangleq \min_{k\in \mathcal{K}} r_k(t).
\end{align}
Maximizing the first utility function corresponds to achieving the highest network throughput, while maximizing the second utility function guarantees a fair user association and improves the throughput for cell-edge UEs who usually suffer from low data rates.
}

Thus, the user association optimization problem at time slot $t$ can be written as
%Thus, the optimization problem for each time slot $t$ can be expressed as
%\mathrm{over}~~&\beta_k(t)\in \mathcal{J}, ~~\forall k \in \mathcal{K}\\
\begin{subequations}\label{opt_prob2}
\begin{align}
\maxi_{\mathbold{\beta}(t)}~~&U(\mathbf{r}(t))\\
\mathrm{subject~to}~~&\sum_{j\in\mathcal{J}} 1_{\beta_k(t)}(j) \leq 1, ~~\forall k\in \mathcal{K}\\
&\sum_{k\in\mathcal{K}}1_{\beta_k(t)}(j). n_k \leq D_j, ~~\forall j\in \mathcal{J}
\end{align}
\end{subequations}
%&\sum_{k\in \mathcal{Q}_j(t)}\textrm{Tr}(\mathbf{F}_{k,j}(t)\mathbf{F}_{k,j}^*(t)) \leq P_j ,~~\forall j\in \mathcal{J}
%%%%%%%%%%%%%%%%%%%%%%%%%%%%%%%%%%%%%%%%%
%This is an optimization problem with the following sets of integer variables: 
This is an optimization problem with integer variables $\beta_k(t)\in\mathcal{J}$.
% for $k\in\mathcal{K}, t\in\mathcal{T}$. 
Here, we use equal power allocation scheme to split the BS power among its active users. Thus, the power constraint in (\ref{power}) is automatically satisfied and can be ignored.
\textcolor{black}{
Note that the set of constraints in (\ref{opt_prob2}c) allows our user association scheme to limit each BS’s load separately, and assuming the number of data streams ($D_j$) is chosen based on the available resources at each BS, this set of constraints guarantees that each BS can serve all its associated UEs simultaneously and thus scheduling is not required. }

As discussed earlier, we adopt the approach of fixing the beamforming scheme and only optimize for user association. For beamforming in the ideal condition of instantaneous CSI and absence of interference, the optimal beamforming vectors of a channel from UE $k$ to a BS $j$ are the eigenvectors corresponding to the first $n_k$ largest singular values, where $n_k$ is the number of data streams intended for UE $k$ \cite{SVD}. This ideal case provides an upper bound on beamforming performance in a realistic network.
%For this case, we need to solve the joint problem of beamforming design and user association which is shown to be NP-hard. 
Considering (\ref{R_kj}), it is clear that our formulation of the user instantaneous rate includes the precoder and combiner matrices of all UEs and BSs which are applicable using any beamforming method. Thus, it is plausible that a user association scheme based on singular value decomposition (SVD) beamforming is compatible with user association based on other beamforming approaches and can achieve near-optimal results.

We use SVD beamforming to obtain the precoder and combiner matrices. To this end, we first need to decompose the channel matrix $\mathbf{H}_{k,j}\in\mathbb{C}^{N_k\times M_j}$ as
\begin{align}
\mathbf{H}_{k,j} &= \mathbf{\Phi}\mathbf{\Sigma}\mathbf{\Gamma}^*
\end{align}
where $\mathbf{\Phi}\in\mathbb{C}^{N_k\times r}$ is the unitary matrix of left singular vectors, $\mathbf{\Sigma}\in\mathbb{C}^{r\times r}$ is the diagonal matrix of singular values (in decreasing order), $\mathbf{\Gamma}\in\mathbb{C}^{M_j\times r}$ is the unitary matrix of right singular vectors, and $r=\textrm{rank}(\mathbf{H}_{k,j})$. Then, we partition the channel matrix as 
\begin{align}\label{Ch_partitioning}
\mathbf{H}_{k,j} &= 
\left[
\begin{matrix}
\mathbf{\Phi}_1 & \mathbf{\Phi}_2
\end{matrix}
\right]
\left[
\begin{matrix}
\mathbf{\Sigma}_1 &  \mathbf{0}\\
\boldsymbol{0} &  \mathbf{\Sigma}_2
\end{matrix}
\right]
\left[
\begin{matrix}
\mathbf{\Gamma}_1^* \\ \mathbf{\Gamma}_2^*
\end{matrix}
\right]
= \mathbf{\Phi}_1\mathbf{\Sigma}_1\mathbf{\Gamma}_1^*+ \mathbf{\Phi}_2\mathbf{\Sigma}_2\mathbf{\Gamma}_2^*
\end{align}
where  $\mathbf{\Phi}_1\in\mathbb{C}^{N_k \times n_k}$, $\mathbf{\Sigma}_1\in\mathbb{C}^{n_k \times n_k}$, and $\mathbf{\Gamma}_1\in\mathbb{C}^{M_j \times n_k}$.
The above partitioning is done to extract the precoder and combiner of appropriate sizes to support the number of data streams of each user $k$. Specifically, each precoder $\mathbf{F}_{k,j}$ and combiner $\mathbf{W}_k$ must be of size $M_j\times n_k$ and $N_k\times n_k$, respectively. 
Then, the SVD precoder and combiner can be obtained as% \cite{SS}
\begin{align}
\mathbf{F}_{k,j}&=\mathbf{\Gamma}_1\\
\mathbf{W}_k&=\mathbf{\Phi}_1.
\end{align}

The optimization problem in (\ref{opt_prob2}) is a highly non-convex MINLP, which is known to be NP-hard due to its non-convex and nonlinear structure and presence of integer variables \cite{MINLP}. The non-convexity and nonlinearity come from the fact that interference structure contains association variables, specifically the indicator function appeared in (\ref{r_k_t}), and constraints (\ref{opt_prob2}b-c). 
\textcolor{black}{
This feature differentiates our formulated optimization problem from user association formulations considered in \cite{Andrews}-\cite{Caire}.}
These MINLP problems are typically difficult to solve due to their combinatorial structure and potential presence of multiple local maxima in the search space. Exhaustive search is the only known technique to guarantee the optimal solution but quickly becomes infeasible with problem size due to an exponential completion time.

Suboptimal and often heuristic methods exist for solving such an MINLP. For example, genetic algorithm (GA) is a such tool which can be effective for solving combinatorial optimization problems.
GA is a method based on natural selection which simulates biological evolution. It iteratively generates and modifies a population of individual solutions. 
After successive generations, the population eventually evolves toward a near-optimal solution \cite{GA}.
In \cite{ICC18}, we used the GA solver provided in Global Optimization Toolbox of MATLAB to solve the optimization problem in (\ref{opt_prob2}). We showed that the GA works effectively for small networks. However, we found that GA does not work well when the network size increases, taking an exceedingly long time to run and still not reaching a good (close to optimal) solution.
In this paper, we propose an efficient algorithm to find a near-optimal solution for our MINLP.

%%%%%%%%%%%%%%%%%%%%%%%%%%%%%%%%%%%%%%%%%%%%%%%%%%%%
%%%%%%%%%%%%%%%%%%%%%%%%%%%%%%%%%%%%%%%%%%%%%%%%%%%%
\section{Worst Connection Swapping (WCS) Algorithm}
In the previous section, we formulated the optimization problem in (\ref{opt_prob2}) which is an NP-hard MINLP. In this section, we introduce an efficient iterative algorithm which converges to a near-optimal solution in a polynomial time. The algorithm is based on the rational that a user association throughput may be suboptimal because of the weakest UE suffering from a weak direct link or high interference from other BSs in the network. 
Thus, swapping the worst UE-BS connection is likely to provide the UE a stronger link (better connection to another BS) and/or reduce the interference, which consequently improves user instantaneous rates and hence the throughput.
This reasoning leads us to design an algorithm to swap the BS in the worst UE-BS connection with all other BSs and then take the best configuration (activation vector) which results in the highest network \textcolor{black}{utility function.} 
%\textcolor{red}{\st{throughput. Since at each iteration we choose the best activation vector among the current and all new activation vectors, the objective function (network sum-rate) is always non-decreasing. }}
Next, we describe the algorithm in detail.

\begin{algorithm}[t]
\SetAlgoLined
%\KwData{LoS, CSI, Distances}
%\KwResult{near-optimal activation vector $\mathbold{\beta}$ }
\textbf{initialization}:\

~~~- Generate an arbitrary feasible activation vector $\mathbold{\beta}^{(1)}$\;
~~~- Find the initial worst connection $(k^{(1)},j^{(1)})=\argminA_{k,j} R_{k,j}(\mathbold{\beta}^{(1)})$\;
\Repeat{stopping criterion achieved}{
\textit{Swap} the $k^{(i)}$th element (activation factor) with all other elements of $\mathbold{\beta}^{(i)}$ to obtain $(K-1)$ new vectors $\mathbold{\beta}^{(i)}_{n}$ with $n\in{\mathcal{K}},~n\neq k^{(i)}$\;
Find the best activation vector $\mathbold{\beta}^{(i+1)}=\mathrm{arg}~\max_{\{\mathbold{\beta}^{(i)}_{n} | n\in{\mathcal{K}},~n\neq k^{(i)}\},~\mathbold{\beta}^{(i)}} ~U(\mathbf{r}(\mathbold{\beta}))$\;
Find the worst connection $(k^{(i+1)},j^{(i+1)})=\argminA_{k,j} R_{k,j}(\mathbold{\beta}^{(i+1)})$\;
\If{$\mathbold{\beta}^{(i+1)}=\mathbold{\beta}^{(i)}$}{
$l^{(m)} \leftarrow \text{mod}(m-1,K)+1$\;
\textit{Switch} the $k^{(i+1)}$th and $l^{(m)}$th elements of $\mathbold{\beta}^{(i+1)}$ to obtain $\mathbold{\beta}^{(i+1)}_\text{sw}$\;
$\mathbold{\beta}^{(i+1)} \leftarrow \mathbold{\beta}^{(i+1)}_\text{sw}$\;
$m \leftarrow m+1$\;
}
\textbf{Stopping criterion}: Break if the best $\mathbold{\beta}$ does not change after $K$ consecutive swapping iterations\;
$i \leftarrow i +1$\;
%\tcp{at the end of matching game, we have $\mathbf{v}_{\textrm{rej}}=\varnothing$ and obtain $\mathbold{\beta}^{n+1}$.}   %% Comment in Algorithm
}
\caption{WCS User Association Algorithm}
\end{algorithm}

\subsection{Algorithm Description}
The proposed algorithm is summarized in Algorithm 1 and works as follows.
It starts with a random  feasible initial activation vector $\mathbold{\beta}^{(1)}$ which satisfies the constraints in (\ref{opt_prob2}b-c), and an initial worst UE-BS connection $k^{(1)}$ obtained from $\mathbold{\beta}^{(1)}$ as described later in (\ref{k^(i+1)}).
The algorithm has two main steps: i) Swapping step, ii) Switching step. In what follows, we describe these steps in detail.
\subsubsection{Swapping step}
At each iteration $i$, we \textit{swap} the worst connection $\mathbold{\beta}^{(i)}(k^{(i)})$, which is the index of the BS with worst UE-BS connection in current activation vector $\mathbold{\beta}^{(i)}$, with all other connections (i.e. other BS entries in the same activation vector $\mathbold{\beta}^{(i)}$). Since the size of an activation vector is $K$, this swapping results in $K-1$ new activation vectors $\mathbold{\beta}^{(i)}_{n}$ with $n\in{\mathcal{K}},~n\neq k^{(i)}$.
Then, the network \textcolor{black}{utility function}, defined in (\ref{u(r(t))})\textcolor{black}{-(\ref{u2(r(t))})}, is computed for the current and new activation vectors, and the one corresponding to the largest network \textcolor{black}{utility function} is chosen as the best activation vector for the next iteration, i.e.,
\begin{equation}
\mathbold{\beta}^{(i+1)}=\mathrm{arg}~\max_{\{\mathbold{\beta}^{(i)}_{n} |n\in{\mathcal{K}},~n\neq k^{(i)}\},~\mathbold{\beta}^{(i)}} ~U(\mathbf{r}(\mathbold{\beta}))
\label{Beta_i+1}
\end{equation}
Next, we find the worst UE-BS connection in $\mathbold{\beta}^{(i+1)}$ by comparing the user instantaneous rates as
\begin{equation}\label{k^(i+1)}
(k^{(i+1)},j^{(i+1)})=\mathrm{arg}~\min_{k,j} R_{k,j}(\mathbold{\beta}^{(i+1)}).
\end{equation}
and then repeat the process until no improvement in the utility function is achieved. Note that the algorithm is applied for any given time slot $t$ to obtain the optimal activation vector $\mathbold{\beta}(t)$. Here, we dropped the time index $t$ for the ease of notation.

%\textcolor{red}{\st{Moreover, we note that the user instantaneous rate $R_{k,j}$ and the network sum-rate. The user instantaneous rate $R_{k,j}$ and the network sum-rate $U(\mathbf{r})$ are functions of the activation vector. Thus, at each iteration, the network sum-rate either remains the same (when $\mathbold{\beta}^{(i+1)}=\mathbold{\beta}^{(i)}$) or improves (when $\mathbold{\beta}^{(i+1)}\neq \mathbold{\beta}^{(i)}$). This guarantees that the objective function is always non-decreasing.}}

\textcolor{black}{
\textit{Proposition 1}: The sequence of network utility function values produced by Algorithm 1 is non-decreasing.}

\textcolor{black}{
\textit{Proof}:
Since the user instantaneous rate $R_{k,j}$ and the network utility function $U(\mathbf{r})$ are both functions of the activation vector, at each iteration based on (\ref{Beta_i+1}), the network utility function either remains the same (when $\mathbold{\beta}^{(i+1)}=\mathbold{\beta}^{(i)}$) or improves (when $\mathbold{\beta}^{(i+1)}\neq \mathbold{\beta}^{(i)}$). 
}

The above procedure works effectively if the objective function is always increasing in all iterations, i.e. $\mathbold{\beta}^{(i+1)}\neq \mathbold{\beta}^{(i)},~\forall i$. 
However, at some point, the current and best activation vectors may be equal. In this case, the algorithm repeatedly produces the same best activation vector (and consequently the same worst connection) in the following iterations. When that happens, swapping does not lead to any more changes in the connections of other users or the network \textcolor{black}{utility function}. Thus, we need to design another step, the switching step, in order to overcome this problem. Note that if $\mathbold{\beta}^{(i+1)}=\mathbold{\beta}^{(i)}$, then the worst connections are equal, i.e. $k^{(i+1)}=k^{(i)}$, but the reverse is not true in general. 
% a criterion to stop the algorithm. Moreover, in this case there 
%If the best connection for a particular user is always the worst connection in the network, the algorithm finds it as the worst connection at each iteration and may output the same best activation vector repeatedly. In this case there is no chance to change the connections of other users and probably improve the network sum-rate. 

\subsubsection{Switching step}
In order to resolve the above issue and find new possibility to improve the objective function, we introduce a switching step. At each iteration, we compare $\mathbold{\beta}^{(i+1)}$ and $\mathbold{\beta}^{(i)}$. If they are not equal, no further action is required and the algorithm proceeds to the next iteration. If they are equal, then we \textit{switch} the worst connection $k^{(i+1)}$ (which is equal to $k^{(i)}$) with another user to find a potentially better activation vector. 
To this end, at each switching step $m$, we compute a UE index as $l^{(m)}=\text{mod}(m-1,K)+1$ and switch the $k^{(i+1)}$th and $l^{(m)}$th elements of the current $\mathbold{\beta}^{(i+1)}$ to obtain the \textit{switched activation vector} $\mathbold{\beta}^{(i+1)}_{\text{sw}}$. Then, we use this vector as the initial activation vector for the next swapping iteration.
Note that the switching process does not affect the non-decreasing nature of the algorithm, since in the swapping step of next iteration, the pre-switching activation vector $\mathbold{\beta}^{(i+1)}$ appears in $\mathbold{\beta}^{(i+2)}_n$ and therefore is included in the \textcolor{black}{utility function} comparisons (see (\ref{Beta_i+1})). 
Moreover, the switching process guarantees that the connection of all users is considered for improvement, since $l^{(m)}$ sweeps the set of all users in order.

The algorithm stops when there is no further improvement in the network \textcolor{black}{utility function} (no change in optimal activation vector) after $K$ consecutive swapping iterations. This criterion is reasonable because $l^{(m)}=l^{(m+K)}$ means after $K$ iterations the algorithm starts to switch the same worst connection with the same user.
We note that the feasibility of the solution is also guaranteed since the algorithm starts with a feasible activation vector and at each iteration just swaps the elements of this vector to find the best activation vector.

\subsection{Complexity Analysis}
In this subsection we \textcolor{black}{consider the sum-rate utility function and} analyze the algorithm complexity by computing the number of floating multiplications. The cost of multiplication of two matrices, $A\in\mathbb{C}^{n\times m}$ and $B\in\mathbb{C}^{m\times p}$, is $\mathcal{O}(nmp)$. SVD decomposition of matrix $A\in\mathbb{C}^{n\times m}$ has a cost of $\mathcal{O}(nm^2)$ ($m\geq n$), matrix inversion by Jordan-Gauss elimination scales as $O(n^3)$, and the cost of computing the matrix determinant is $\mathcal{O}(n^3)$. 
Finding $\mathrm{arg}~\max$ or $\mathrm{arg}~\min$ over a set of $n$ variables, also known as sorting algorithms, requires $n\log(n)$ comparisons \cite{IntroAlg}.
%https://en.wikipedia.org/wiki/Sorting_algorithm
When computing the multiplication of three matrices $A\in\mathbb{C}^{n\times m}$, $B\in\mathbb{C}^{m\times p}$, and $C\in\mathbb{C}^{p\times q}$, the order in which the product is parenthesized is important, and we take this into account to minimize the cost \cite{Boyd}.
Based on these complexity orders, we can analyze the complexity of our WCS algorithm. We first look at the complexity order of a single swapping iteration, then examine the number of iterations. At each swapping iteration of the algorithm, Steps 6 and 7 have the highest computational cost because of the search/sort procedures, and according to above complexity bases, they exhibit similar computation order. Thus, we focus on computing the cost of Step 6 which involves two major operations: computing the network sum-rate \textcolor{black}{$U_s(\mathbf{r}(t))$}, and sorting the resulted sum-rate vector. Computing the network sum-rate involves computing the interference plus noise term $\mathbf{V}_{k,j}$ in (\ref{Y_interf}) which is of order $\mathcal{O}(M_j^2)$, and this leads to the cost of computing user instantaneous rate $R_{k,j}$ in (\ref{R_kj}) also as $\mathcal{O}(M_j^2)$. Thus, the cost of computing the network sum-rate in (\ref{u(r(t))}) is of the order $\mathcal{O}(M_j^2)$. 
Next, the cost of sorting and finding $\mathrm{arg}~\max$ over the set of $K-1$ new activation vectors $\mathbold{\beta}_n^{(i)}$ and the previous activation vector $\mathbold{\beta}^{(i)}$ is $\mathcal{O}(K\log(K))$. Therefore, the total cost of computing Step 6 is $\mathcal{O}(K\log(K)M_j^2)$, and since Step 6 is computationally dominant in each iteration, this is also the order of the cost for a single swapping iteration. 
Based on the algorithm's stopping criterion, the number of swapping iterations is between $K$ and $cK$ iterations with $c$ as a constant factor, $1\leq c\ll K$, and we can conclude that the number of iterations increases linearly with $K$. Thus, the final complexity of the algorithm is $\mathcal{O}(K^2\log(K)M_j^2)$.
 
\section{Numerical Results}\label{Sim_sec}
In this section, we evaluate the performance of the proposed user association scheme in the downlink of a mmWave MIMO network with $J$ BSs and $K$ UEs operating at 73 GHz. \textcolor{black}{In our simulations, we consider the network sum-rate utility function defined in (\ref{u(r(t))}), except when study user fairness and cell-edge users' rate in Section \ref{Sim_sec}.A.5.} The mmWave links are generated as described in Section II-A, each link is composed of 5 clusters with 10 rays per cluster.
Each BS is equipped with a $8\times 8$ UPA antenna, and \textcolor{black}{each UE is equipped with a $2\times 2$ mmWave UPA antenna for 5G connections}. The noise power spectral density is $-174$ dBm/Hz. \textcolor{black}{Unless otherwise stated,} all BSs transmit at the same power level $P_j$.
Moreover, we assume that the network nodes are deployed in a region of size $300~\textrm{m} \times 300~\textrm{m}$. The BSs are placed at specific locations and the UEs are distributed randomly within the given area. There are $n_k$ data streams for each UE and $Q_j(t)$ is the maximum number of allowed active users at each BS. Thus, the total number of data streams sent by each BS is $D_j=n_kQ_j(t)$. A BS is considered to be overloaded (congested) if more than $Q_j(t)$ UEs are associated with that BS. 
%Throughput values are obtained by averaging over ******** channels at ********* random UE locations.

We present our numerical results in two main thrusts. First is the performance of the proposed user association scheme for mmWave with respect to its novel features: (i) using both instantaneous and large-scale CSI, (ii) considering the dependency of interference on user association, and (iii) using MIMO beamforming at both ends. Second is the numerical analysis of our proposed WCS algorithm in terms of convergence and complexity to verify the derived theoretical complexity order.

\subsection{Performance of the proposed user association scheme}
\begin{figure}%
\centering
\subfigure[]{%
%\label{fig:first}%
\includegraphics[height=2.5in]{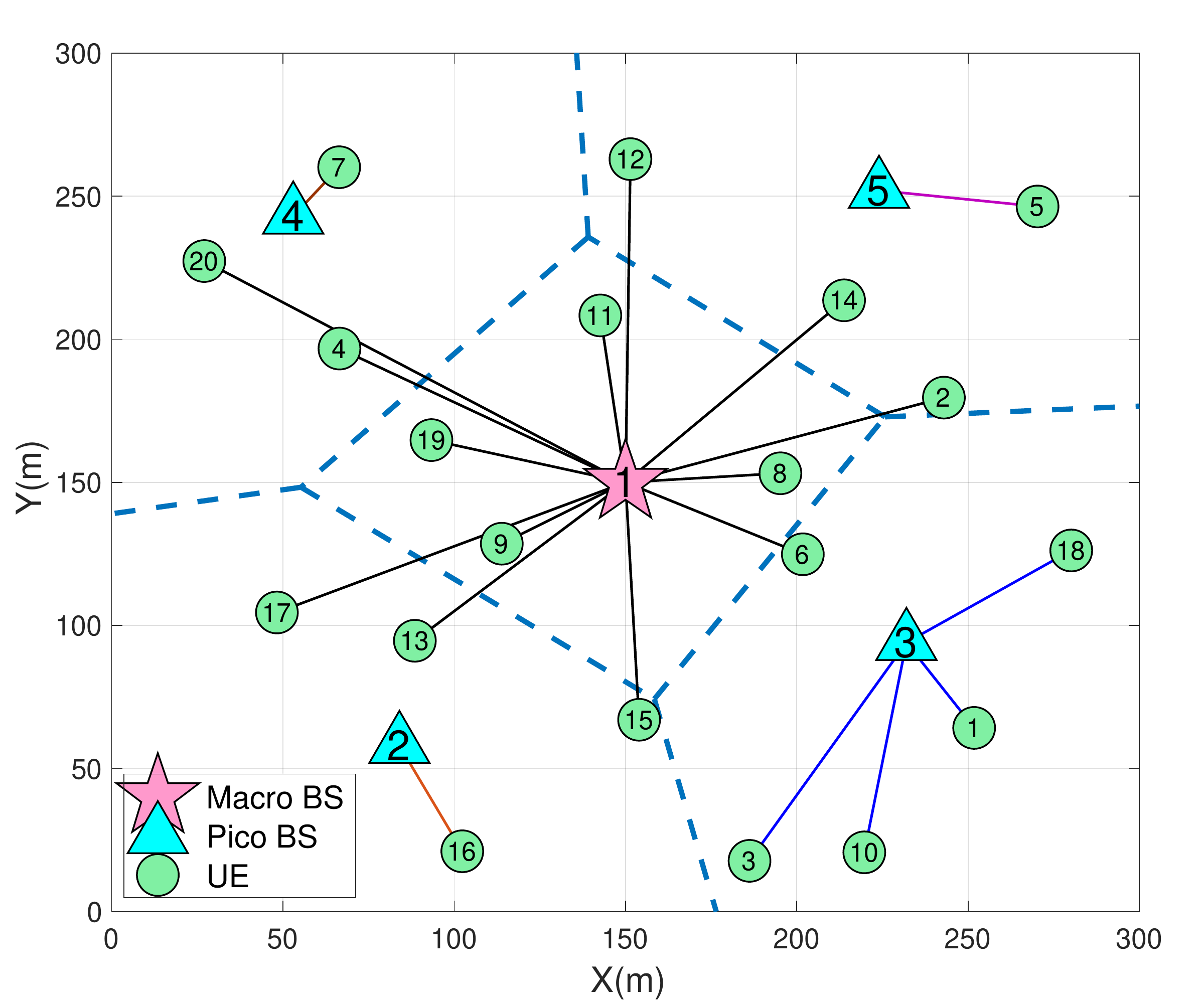}}%
\qquad
\subfigure[]{%
%\label{fig:second}%
\includegraphics[height=2.5in]{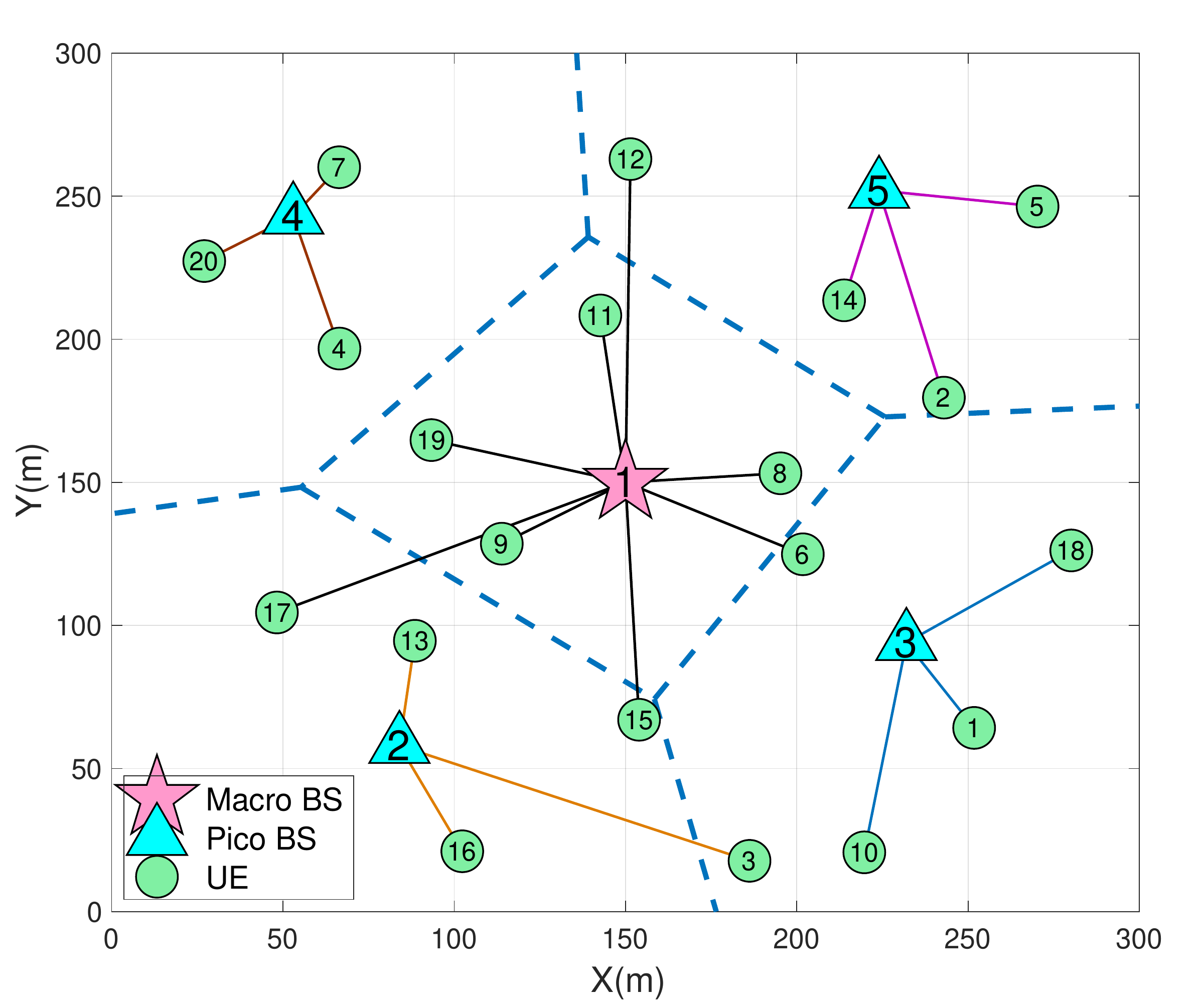}}%
\vspace*{-1em}
\caption{Comparison of user assocaition schemes in a HetNet with 1 Macro BS and 4 Pico BSs. (a) Max-SINR user association with full interference, (b) Proposed load balancing user association.}
\vspace*{-1em}
\label{UA_figs}
\end{figure}
\subsubsection{Comparison of max-SINR and proposed load balancing user association schemes}
\textcolor{black}{
First, we compare the conventional max-SINR user association scheme with our proposed load balancing user association scheme. 
In order to better reflect the capabilities of the proposed user association scheme, in this simulation study, we consider a two-tier HetNet with 1 macro BS (centered at the middle of the network), 3 pico BSs and 20 UEs. The transmit power of macro BS is 30 dBm, while pico BSs transmit at 20 dBm. 
\textcolor{black}{The microwave links are i.i.d. Rayleigh fading MIMO channels, and we assume that each UE is also equipped with a single-antenna module designed for LTE/4G connections at microwave band in addition to their mmWave antennas}. 
The macro BS can accommodate 8 UEs, and each pico BS can only serve 3 UEs.
Fig. \ref{UA_figs}.a shows the result of max-SINR association in the network. As we can see from the figure, the macro BS is overloaded by 4 extra UEs, and BS \#3 is overloaded with one extra UE. 
%So, these UEs need be pushed into other BSs with available resources (data streams). 
User association using our proposed scheme is shown in Fig. \ref{UA_figs}.b. It can be seen from the figure that the proposed method perfectly balances the BSs' load by pushing the overloading UEs from the congested BSs into the lightly-loaded BSs. 
}
\begin{figure}
\centering
\includegraphics[height=2.9in]{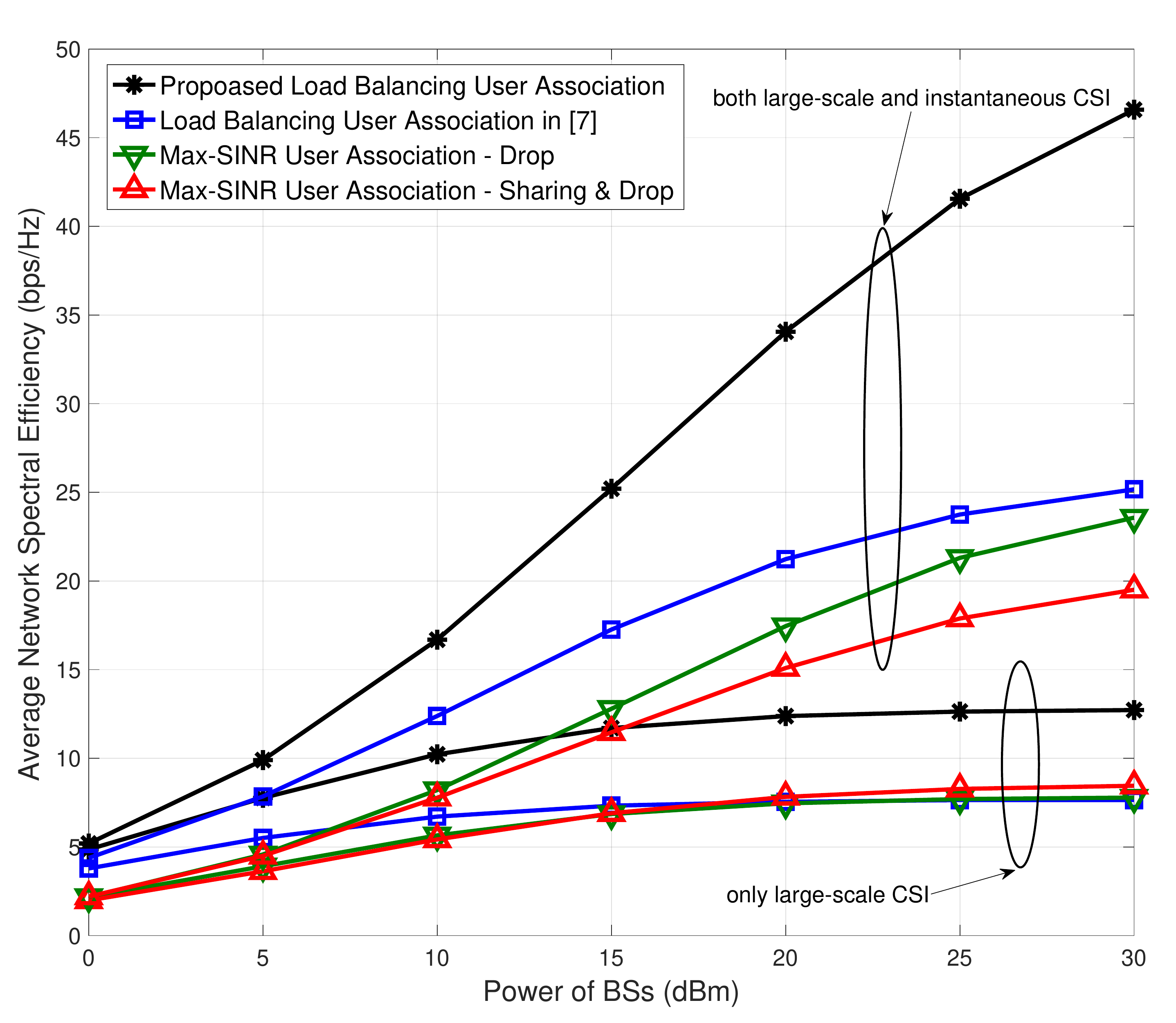}
\caption{Average network spectral efficiency for user association schemes in a mmWave network with $3$ BSs, $12$ UEs, $8\times8$ UPA at each BS, and $2\times2$ UPA at each UE. The UEs are distributed randomly in an area of size $300~m\times300~m$. All BSs transmit at the same power level, and the noise power spectral density is $-174$ dBm/Hz.}
\label{SumRate}
\end{figure}
\subsubsection{Effect of network interference dependency on user association}
\textcolor{black}{
Next, we compare throughput performance of the proposed user association method by comparing it with three other user association schemes: (i) load balancing user association \cite{Caire}, (ii) max-SINR user association with user drop, and (iii) max-SINR user association with resource (data stream) sharing and user drop. The first scheme is a load balancing user association which results in fractional association coefficients, because the integer associaiton constraint is relaxed to solve the optimization poblem \cite{Caire}. In the second method, those UEs who overloaded the congested BS are dropped. For the third scheme, the data streams of the congested BS are shared among the maximum number of UEs it can serve, and subsequent UE connections are dropped. For example, consider a network with 4 BSs and 8 UEs where $n_k=2,~\forall k$, and $D_j=4,~\forall j$. If a BS is overloaded with 5 UEs, it shares its 4 data streams among 4 UEs (reducing the number of data stream for each UE to 1) and drops the remaining UE.
In all these three reference schemes, the interference is assumed to be both independent of user association and present all the time (full interference), while our proposed scheme considers the dependency between the network interference structure and user association.
%For the last scheme, we perform the load balancing user association based on the approach presented in \cite{Caire}, which ignores the effects of instantaneous CSI on user association and assumes that the interference from other BSs is independent of user association and present all the time.
}

%Fig. \ref{SumRate} depicts the network spectral efficiency (given in (\ref{u(r(t))})) versus the BSs' transmit power for these different association schemes. 
\textcolor{black}{
Fig. \ref{SumRate} compares the network spectral efficiency (given in (\ref{u(r(t))})) for the user association schemes with two different CSI assumptions: (i) both large-scale and instantaneous CSI, and (ii) only large-scale CSI. As we can see from the figure, our proposed scheme shows a better performance in both cases since we adapt the network interference to the association.
}
Thus, It can be inferred from the figure that user association is highly dependent on network interference, since our proposed method outperforms the other user association schemes which all ignore the effect of user association on the network interference. Further, the availability of instantaneous CSI in mmWave channels can dramatically boost performance, and this effect is observed across all schemes, though most pronounced in our proposed scheme.

%Also, we can see that the load balancing scheme presented in \cite{Caire} slightly underperforms the max-SINR schemes. This result makes sense since, contrary to max-SINR schemes, the load balancing approach takes into account the BS loads.

\begin{figure}
\centering
\includegraphics[height=2.9in]{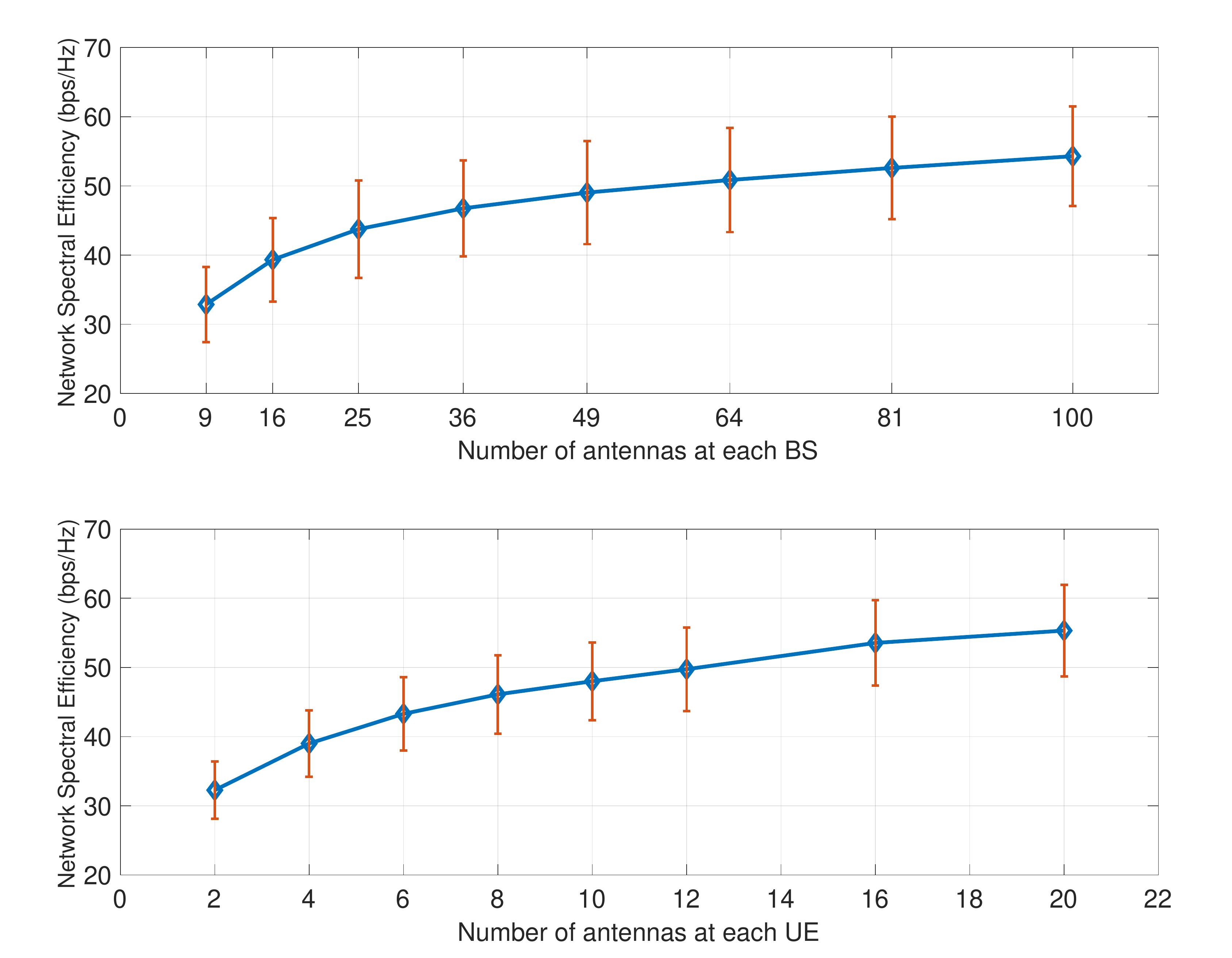}
\caption{Network spectral efficiency versus the number of antennas at each BS (upper subfigure) and at each UE (lower subfigure).}
\label{SS_Mj_Nk}
\end{figure}

%\begin{figure}
%\centering
%\includegraphics[scale=.45]{M77_Alg4_SS_vs_Nk.pdf}
%\caption{Network spectral efficiency versus the number of antennas at each UE.}
%\label{SS_Nk}
%\end{figure}

\subsubsection{Effect of antenna array size on network spectral efficiency}
Fig. \ref{SS_Mj_Nk} depicts the network spectral efficiency versus the number of antennas at each BS and each UE. There are 12 UEs and 4 BSs in the network and the number of data stream at each UE is $n_k=2$. The average is taken over 500 mmWave channel realizations, and the error bars represent the standard deviation. In the upper subfigure, we fix the number of antennas at each UE ($N_k=4$), and increase the antenna array size at each BS. The result shows that increasing the antenna array size from a $3\times 3$ UPA to a $10\times 10$ UPA improves the network spectral efficiency by 164\%. Similarly, in the lower subfigure we increase the antenna array size at each UE while fixing the number of antennas at each BS at $M_j=64$. The simulated set of UPA at each UE is: $\{1\times 2,~2\times 2,~2\times 3,~2\times 4,~2\times 5,~3\times 4,~4\times 4,~5\times 4\}$. The results show a 170\% improvement in the network spectral efficiency when we increase the UE antenna array size from $1\times 2$ to $5\times 4$. Both results show that increasing the number of antennas at either the UE or BS has a positive impact on the network spectral efficiency, although the improvement is more significant at small numbers of antennas.

\begin{figure}
\centering
\includegraphics[height=2.9in]{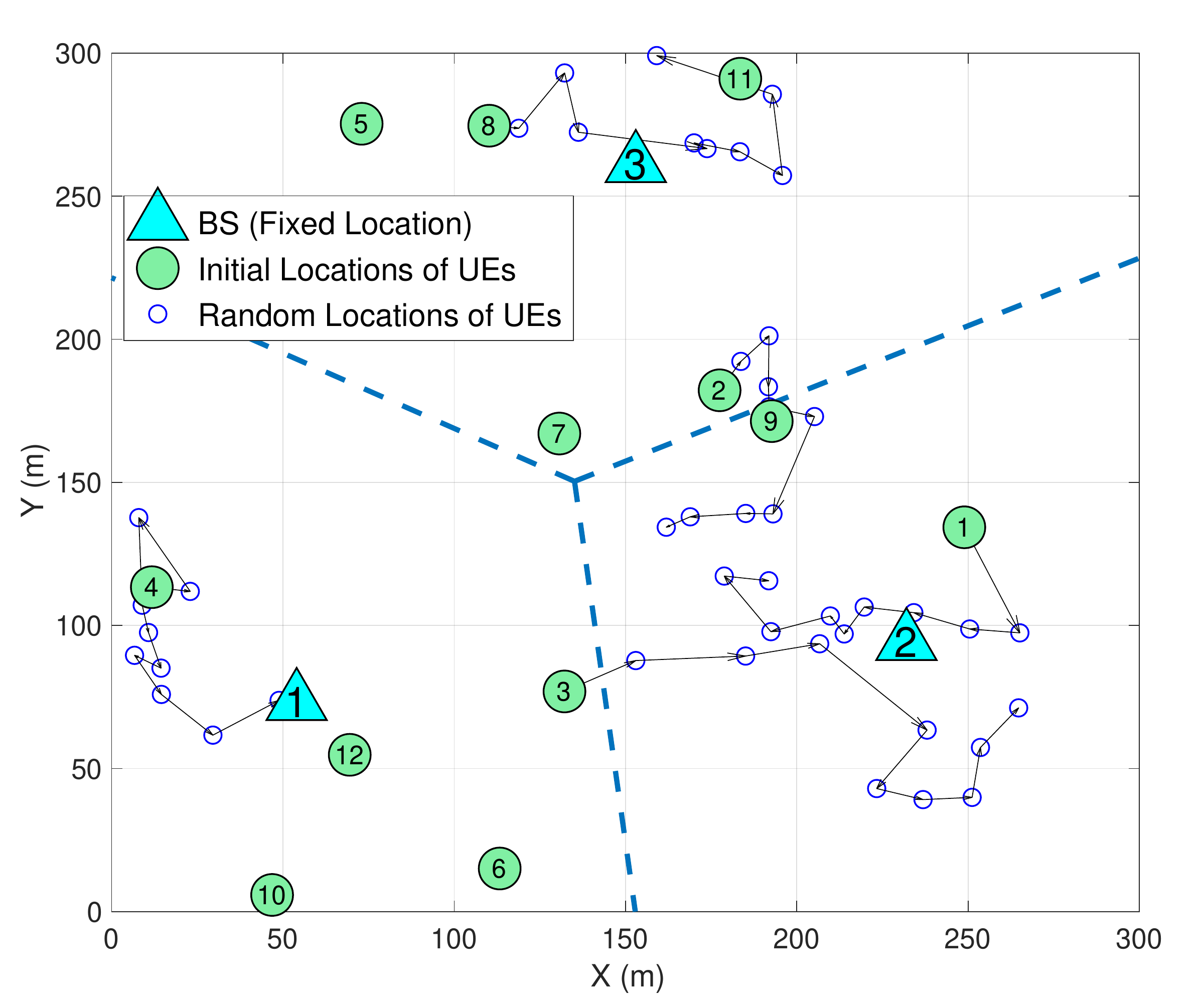}
\caption{A mmWave network with 3 BSs and 12 UEs. All UEs move randomly from one time slot to another. The trajectory is shown only for couple of UEs.}
\label{Net_top_Mobility}
\end{figure}

\begin{figure}%
\vspace*{-1em}
\centering
\subfigure[]{%
%\label{fig:first}%
\includegraphics[height=2.25in]{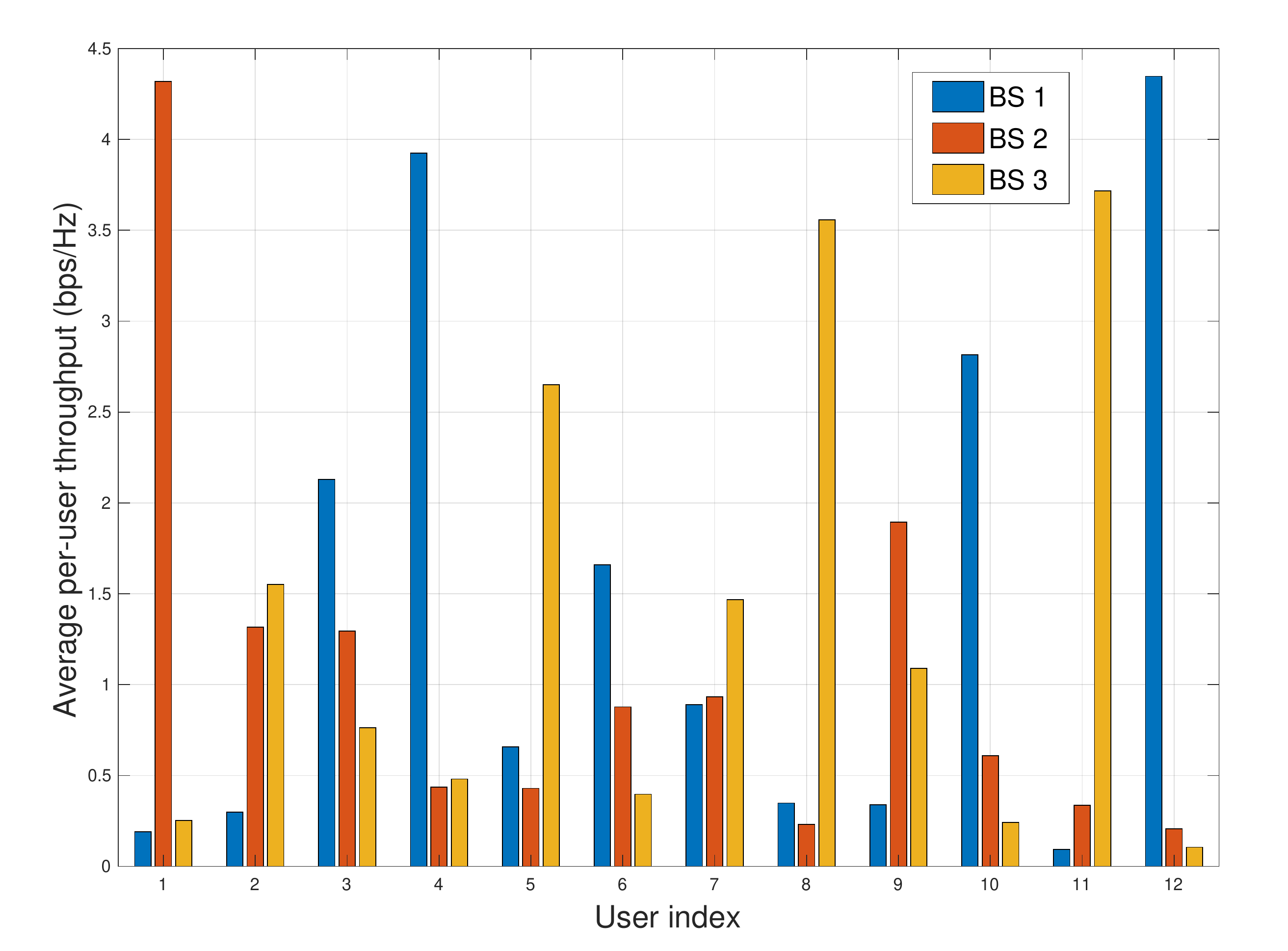}}%
\qquad
\subfigure[]{%
%\label{fig:second}%
\includegraphics[height=2.25in]{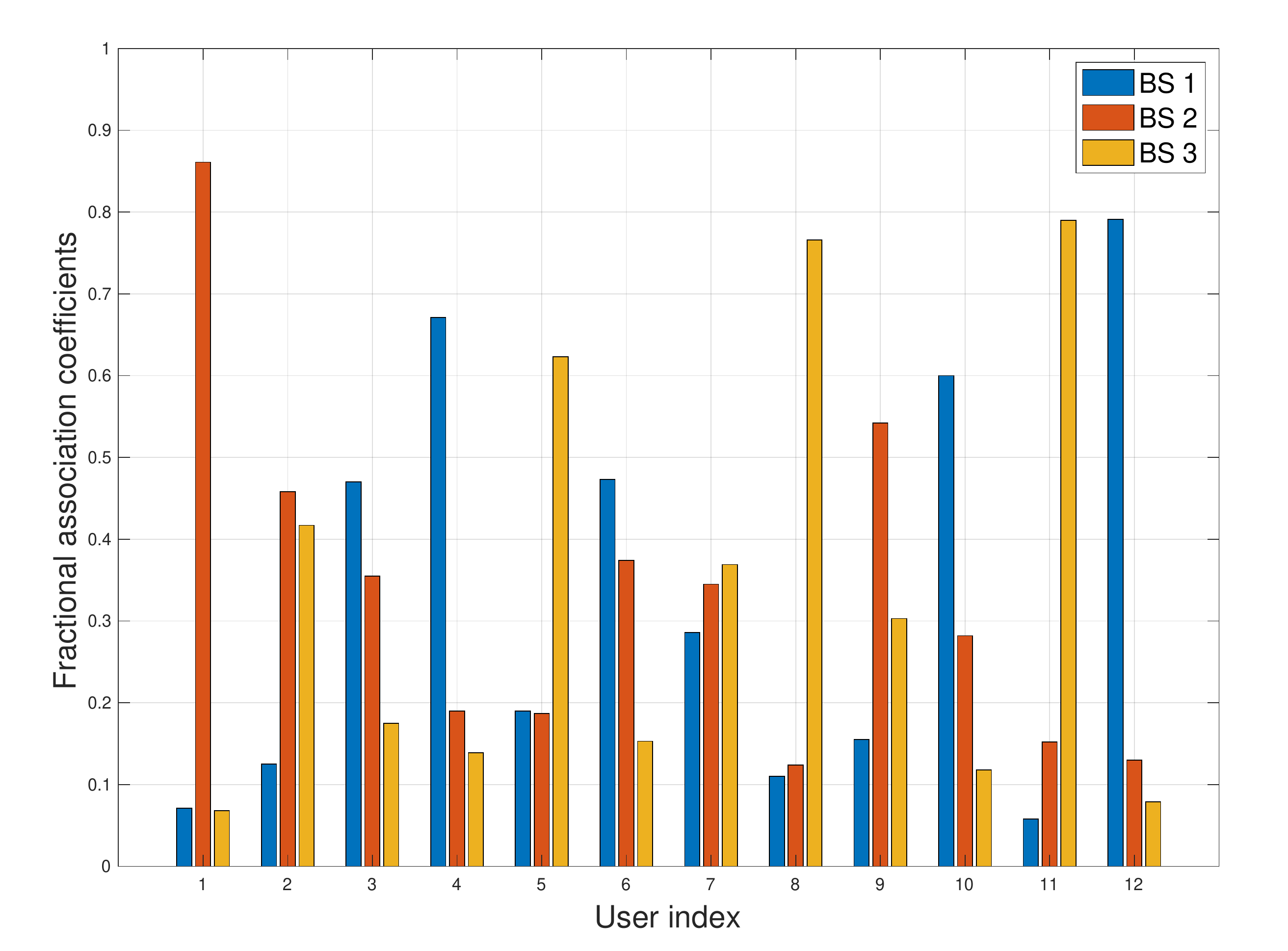}}%
\caption{(a) Average per-user throughput and (b) fractional association coefficients for the mmWave network depicted in Fig. \ref{Net_top_Mobility}.}
\label{Ass_coeff_Mobility}
\end{figure}

\textcolor{black}{
\subsubsection{Effect of mobility on user association}
We consider a simple mobility model where the location of UEs are fixed during each time slot. In the subsequent slot, each UE moves to a new random location within a 5m x 5m square box centered at its location in the previous time slot. The user association is performed at each time slot and the fractional user association is obtained by taking average over $T=1000$ time slots. 
Fig. \ref{Net_top_Mobility} depicts the network topology with random locations of UEs shown in small circles. Note that the number of random locations for each UE is 1000, but we only show 10 random locations ($t=1:100:1000$) for each UE to avoid a messy figure. Fig. \ref{Ass_coeff_Mobility} depicts the average per-user throughputs and the fractional associations. 
As we can see from these simulations, UE 4 is always moving around BS 1 and thus it is associated with BS 1 for about 87 percent of the time. The same result can be inferred for UE 1 which is connected to BS 2 for most of the time. UE 2, however, gradually moves away from BS 3 and gets closer to BS 2, and that is the reason this UE is associated with both BSs for about half of the time. UE 2 therefore could experience frequent handover in this case. These simulations confirm that our proposed user association is well-designed to be performed per-time slot, and it can be applied with any mobility model.
}

\begin{figure}%
\centering
\subfigure[]{%
%\label{fig:first}%
\includegraphics[height=2.5in]{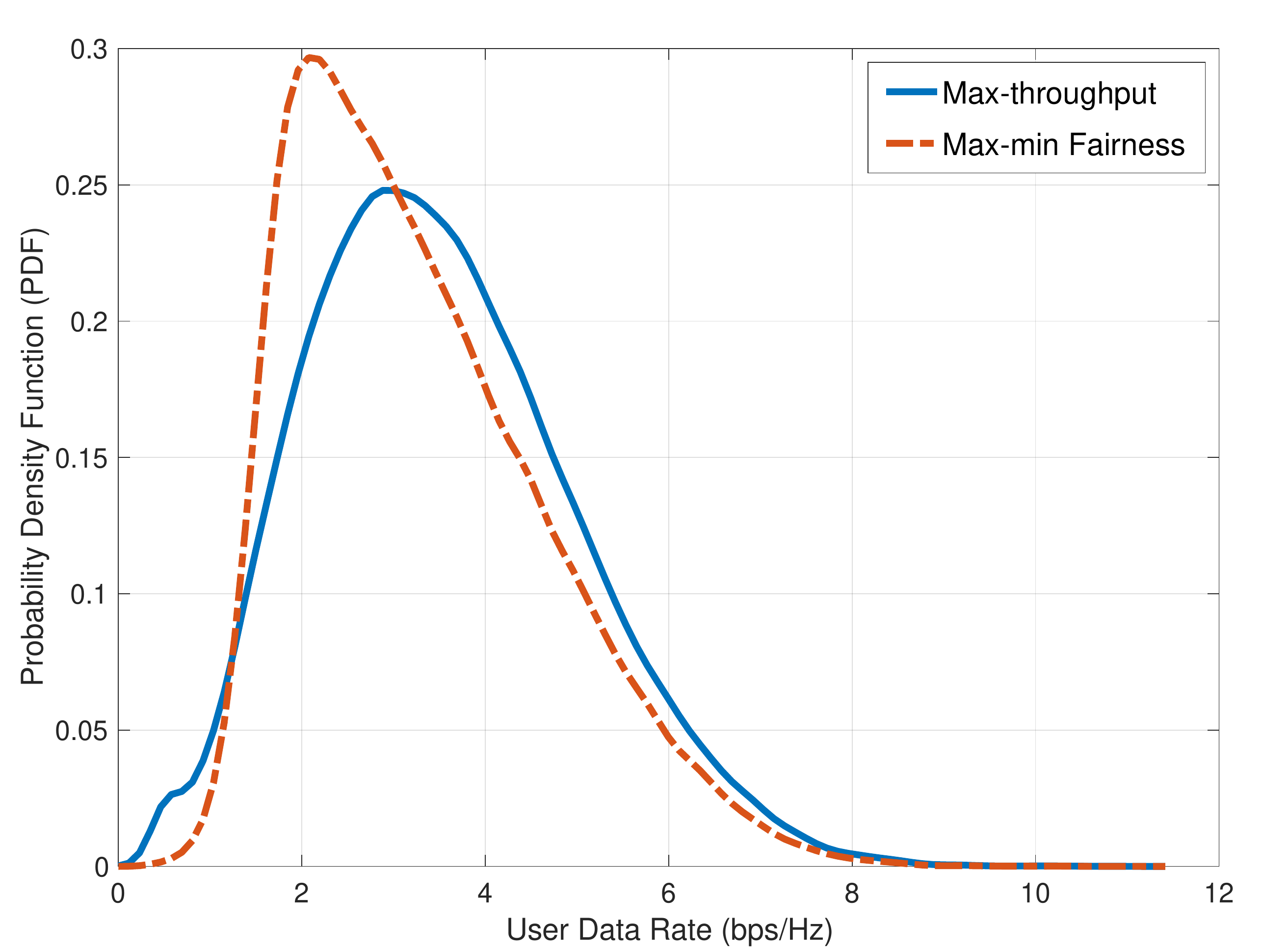}}%
\qquad
\subfigure[]{%
%\label{fig:second}%
\includegraphics[height=2.5in]{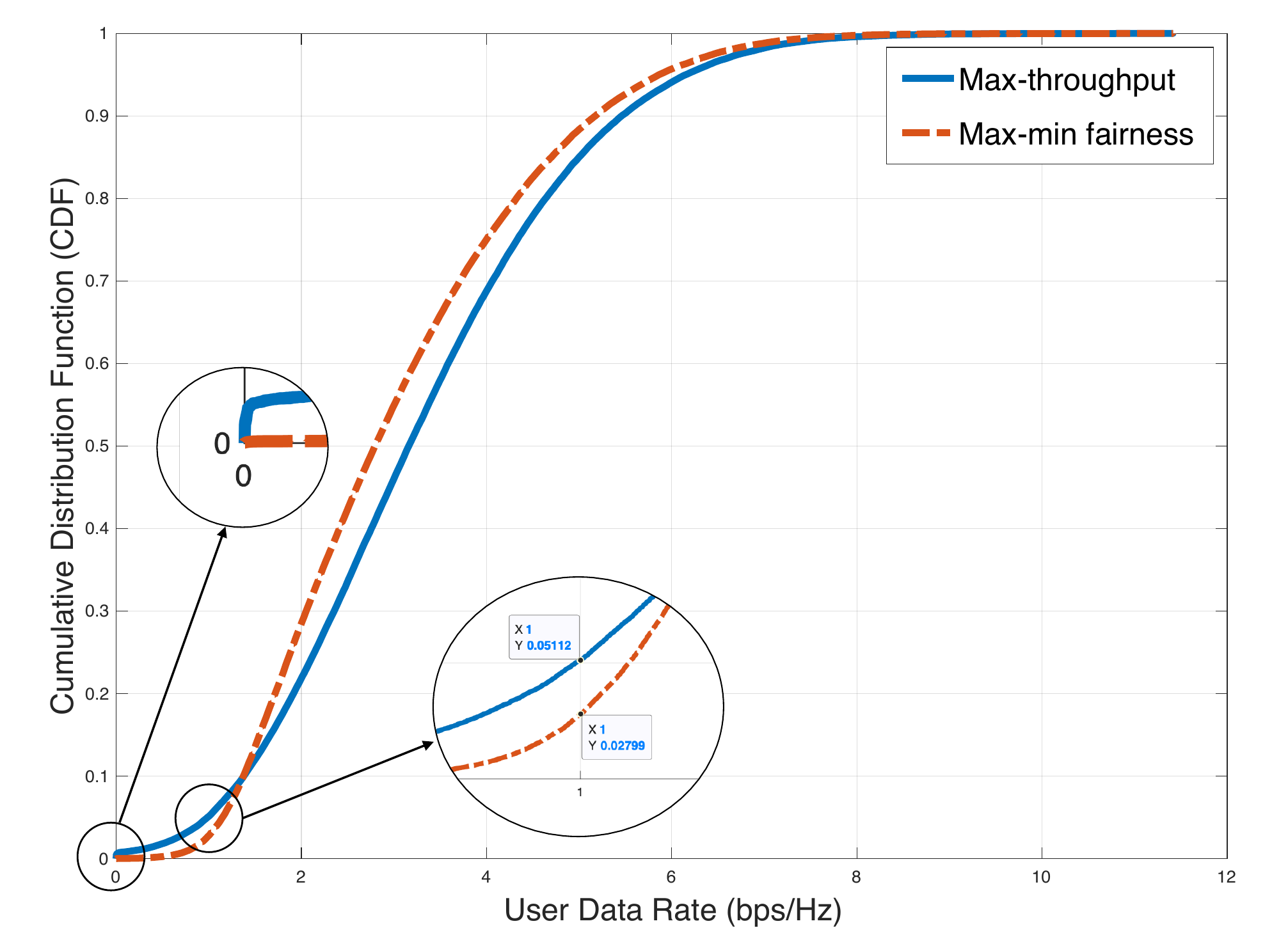}}%
\caption{Comparing the effect of fairness in a mmWave network with 3 BSs and 15 UEs. (a) PDF, and (b) CDF of users' data rates.}
\label{Fairness}
\end{figure}
\textcolor{black}{
\subsubsection{Effect of fairness on cell-edge users' throughputs}
Cell-edge users usually suffer the worst throughput in the network. In order to study the the effect of fairness on cell-edge throughput, we compare the two utility functions defined in (\ref{u(r(t))})-(\ref{u2(r(t))}).
Fig. \ref{Fairness} shows the probability density function (PDF) and cumulative distribution function (CDF) of users' data rates in a mmWave network with 3 BSs and 15 UEs. We consider the sum-rate utility function in (\ref{u(r(t))}) for the ``Max-throughput'' case, and use the min-rate utility function  in (\ref{u2(r(t))}) to implement “Max-min Fairness”.
By focusing on the low rate region (bottom left corner) of each figures, it can be inferred that compared to max-throughput, max-min fairness shows a lower probability of users having an extremely low (close to zero) rate. For example, in Fig. \ref{Fairness}.b the probability of having users with data rate smaller than 1 bps/Hz is 5.1\% for sum-rate utility function, while this probability is about 2.8\% when using the max-min fairness. Since most cell-edge users would be represented in the low-rate region, this result implies that max-min fairness results in higher data rates for the cell-edge users in the network, especially ensuring that these users' rates start at a reasonable non-zero value. This is in contrast to max-throughput in which the CDF figure shows a small portion of users actually having zero rates. This result confirms that max-min fairness improves the throughput for cell-edge users who usually suffer from low data rates. 
}

% can use the CDF to say the probability of having a rate smaller than something then can read it off the CDF curve itself

\subsection{Efficiency of the proposed WCS algorithm}
\subsubsection{Convergence performance}
\begin{figure}
\centering
\includegraphics[scale=.32]{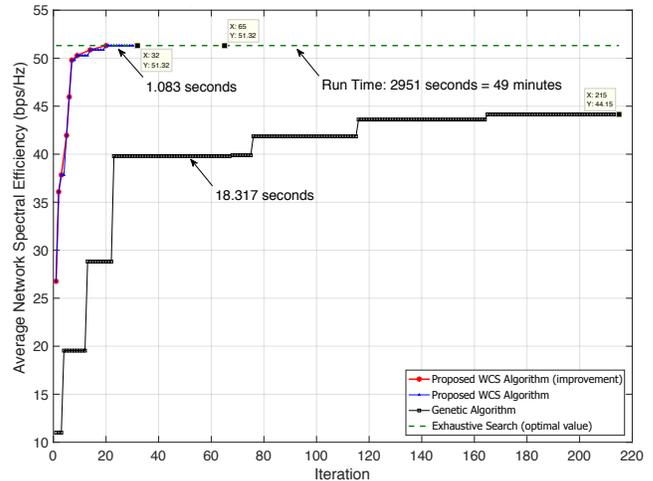}
\caption{Comparison of performance of different algorithms in a network with 4 BSs and 12 UEs. GA is implemented with an infinite time limit and 50 stall generations.}
\label{Alg_vs_GA_vs_ES}
\end{figure}

Fig. \ref{Alg_vs_GA_vs_ES} compares the performance of the proposed algorithm with GA in a network with 12 UEs and 4 BSs. As a benchmark, exhaustive search result is included to provide the optimal solution.
The ``\textit{improvement}" curve shows the increment of network sum-rate at each swapping iteration. It can be seen from the figure that the proposed WCS algorithm rapidly converges to the optimal solution in 32 iterations which execute in 1.08 seconds. However, the GA (with 50 stall generations) stops after 215 iterations and still does not reach the optimal solution. GA can, in theory, reach to the optimal solution by increasing the number of stall generations, but it takes drastically more time to reach this. Our simulations show that it takes 417 iterations which is about 86 seconds for GA (with 200 stall generations) to reach the optimal solution, compared to the proposed WCS which takes only 1.08 seconds. Note that the GA stops if there is no improvement in the objective function for a sequence of consecutive generations of the length equal to the number of stall generations.

\begin{figure}%
\centering
\subfigure[]{%
%\label{fig:first}%
\includegraphics[height=2.7in]{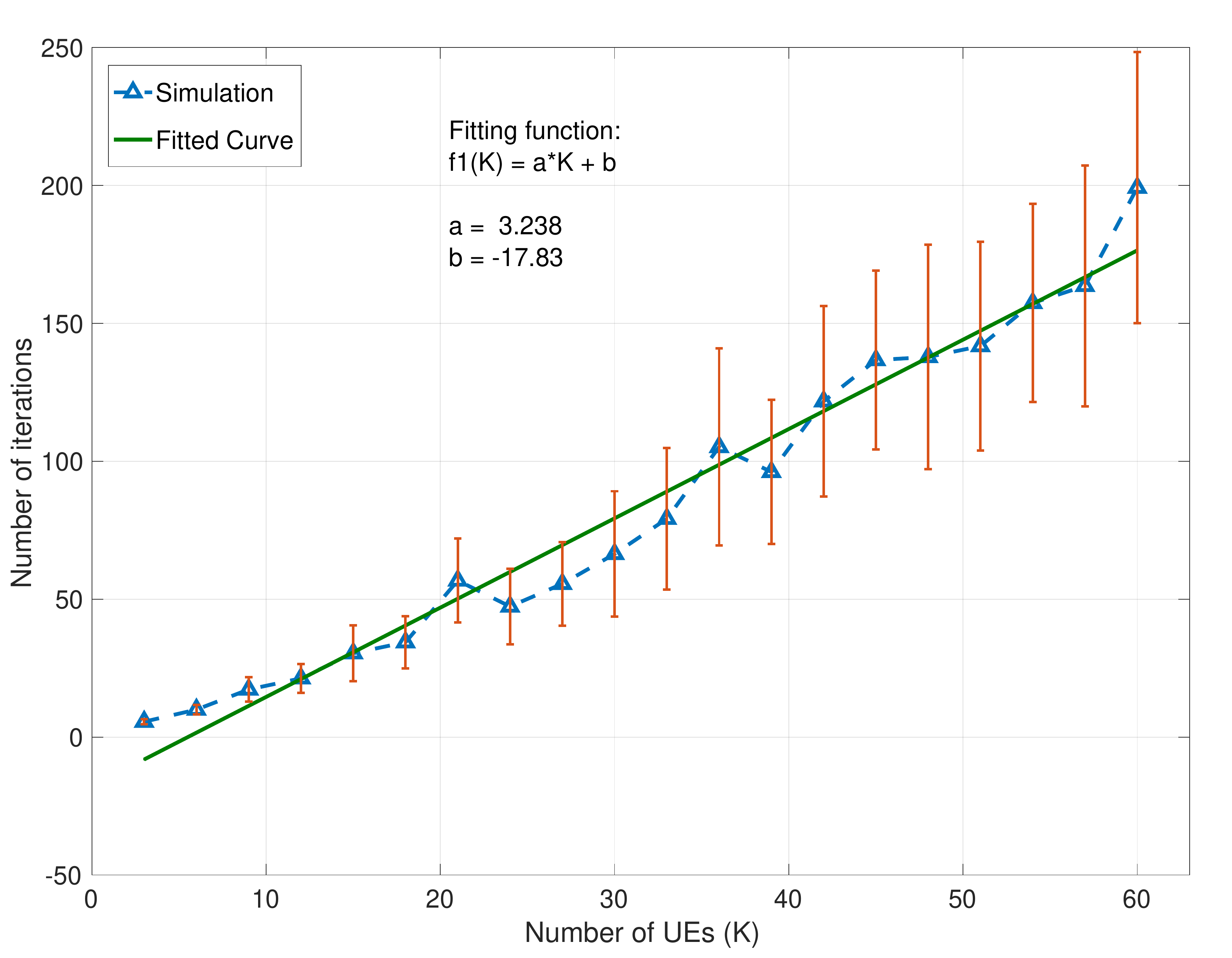}}%
\qquad
\subfigure[]{%
%\label{fig:second}%
\includegraphics[height=2.7in]{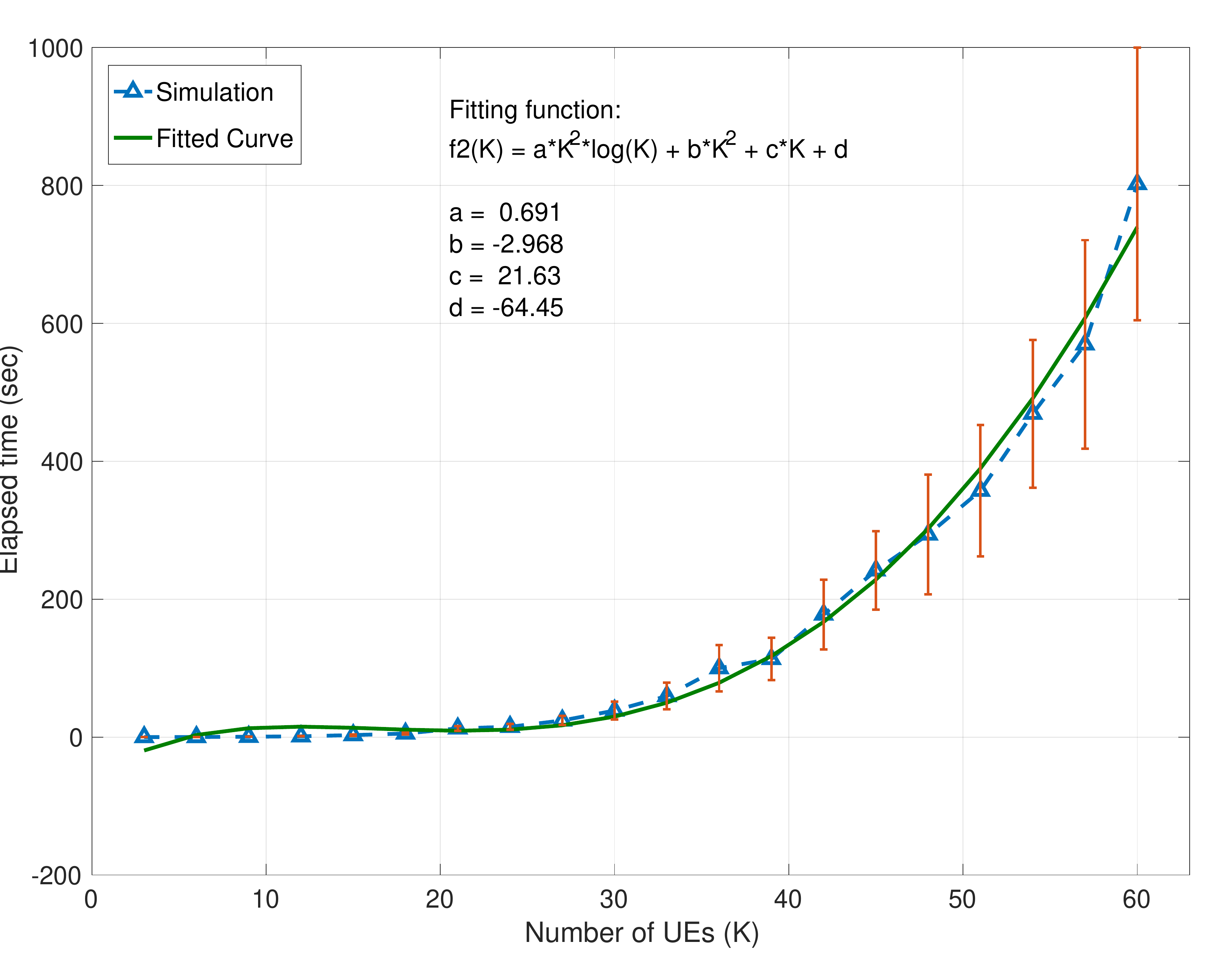}}%
\caption{(a) Number of iterations and (b) elapsed time of the proposed algorithm in a network with 3 BSs and $K$ UEs. The error bars show the standard deviation.}
\label{Fitting_figs}
\end{figure}

\subsubsection{Complexity results}
Comparing the convergence time of the proposed algorithm and GA in Fig. \ref{Alg_vs_GA_vs_ES}, one can easily confirm that the WCS algorithm has much lower complexity than GA.
Fig. \ref{Fitting_figs}.a shows the number of swapping iterations of the proposed algorithm versus the number of UEs while the number of BSs is fixed ($J=3$). The average is taken over 500 channel realizations and polynomial data fitting is implemented. As expected, there is a linear relationship between the number of iterations and the number of UEs. 
Fig. \ref{Fitting_figs}.b depicts the actual elapsed time of the proposed algorithm.
Based on our complexity analysis in Section V-B, we fit the elapsed time results with a function of the form $f(K)=aK^2\log(K)+bK^2+cK+d$, where $K$ is the number of UEs and $a$, $b$, $c$, and $d$ are some constants. The result confirms the complexity analysis and shows that the fitting model matches well with the simulation results. 
%We note that extremely similar polynomial fittings are also obtained for a different number of BSs but do not include the results here for clarity of the plots.

\section{Conclusion}
In this paper we investigated the problem of optimal user association in a mmWave MIMO network. 
We first introduced the activation matrix and showed that the user instantaneous rate is a function of the activation factors.
Then, we formulated a new association scheme in which network interference depends on user association. We designed an efficient polynomial-time algorithm to solve the formulated MINLP optimization problem.
Simulation results showed that the proposed user association scheme outperforms existing association methods, which ignore the effect of interference on user association, confirming the fact that the network interference structure is highly dependent on user association.
Moreover, we designed an efficient algorithm to solve the formulated MINLP optimization problem.
Numerical results confirmed the polynomial complexity of the proposed algorithm and showed that it has significantly lower complexity than the genetic algorithm, and reaches a near-optimal solution in a reasonable time and number of iterations.
%***********************************************************************
%\vspace*{1pt}

\end{document}